# Transmission of hydrogen detonation across a curtain of dilute inert particles

Yong Xu[1], Pikai Zhang[1,2], Qingyang Meng[2], Shangpeng Li[1], and Huangwei Zhang[1,†]

[1] *Department of Mechanical Engineering, National University of Singapore, 9 Engineering Drive 1, Singapore, 117576, Republic of Singapore*
[2] *National University of Singapore (Chongqing) Research Institute, Liangjiang New Area, Chongqing 401123, People's Republic of China*

**Abstract**

Transmission of hydrogen detonation wave (DW) in an inert particle curtain is simulated using the Eulerian─Lagrangian approach with gas-particle two-way coupling. A detailed chemical mechanism is used for hydrogen detonative combustion and parametric studies are conducted based on a two-dimensional computational domain. A detonation map of propagation and extinction corresponding to various particle sizes, concentrations, and curtain thicknesses is plotted. It is shown that the critical curtain thickness decreases considerably when the particle concentration is less than the critical value. The effects of curtain thickness on the trajectories of peak pressure, shock front speed, and heat release rate are examined. Three propagation modes of the DW in particle curtain are found: detonation transmission, partial extinction and detonation re-initiation, and detonation extinction. The chemical explosive mode analysis confirms that a detonation re-initiation event is caused by a re-initiation point with high pressure and explosive propensity, resulting from transverse shock focusing. The influence of particle dimeter/concentration, and curtain thickness on the DW are also examined with peak pressure trajectories, shock speed, and interphase exchange rates of energy and momentum. Furthermore, the evolutions of curtain morphologies are analyzed by the particle velocity, volume fraction, Stokes drag and Archimedes force. This analysis confirms the importance of the drag force in influencing the change of curtain morphologies. Different curtain evolution regimes are found: quasi-stationary regime, shrinkage regime, constant-thickness regime, and expansion regime. Finally, the influences of the curtain thickness on the characteristic time of curtain evolutions are studied.

**Key words:** Hydrogen; detonation re-initiation; detonation extinction; inert particle; particle concentration; curtain thickness

# 1. Introduction

Hydrogen is a promising fuel for decarbonization in energy sectors. Compared to other fuels (e.g., methane), hydrogen has lower ignition energy and wider flammability limit (Olmos & Manousiouthakis 2013), and hence is prone to ignition and explosion. Therefore, safety measures to inhibit the accidental explosion of hydrogen should be carefully implemented and evaluated. Chemically inert fine particles are one of the promising explosion inhibitors considering that they can be readily obtained, and are also cheap and safe to be used without additional damage (Olmos & Manousiouthakis 2013; Fomin & Chen 2009; Liu, et al. 2013). According to the literature (Fomin & Chen 2009; Liu, et al. 2013), by implementing a particle curtain (Fedorov, Tropin & Bedarev 2010; Tropin & Fedorov 2014), the overpressure, propagation speed, and product gas temperature of detonation and/or blast waves can be effectively reduced, thereby minimizing the damage to surrounding infrastructure and personnel.

The influence of inert particle curtains on the detonation wave (DW) has been extensively studied. Kratova and Fedorov (2014) found that the detonation speed is reduced in a two-phase oxygen/non-reactive and aluminum particles mixture. Meanwhile, Khmel and Fedorov (2014) observed that particle collisions have negligible effects on the detonation speed, cell size, and gas parameters. Liu et al. (2016) confirmed that curtain concentration, particle material density, and particle size play key roles in detonation inhibition. In addition to the above studies, Tahsini (2016) found that the residence time of detonation wave in the curtain significantly affects detonation suppression.

Fedorov and Kratova (2015a, 2015b) observed that detonation propagation is more significantly influenced by the composition and distribution of the particles than by particle size and volume fraction. Fedorov and his co-authors (2010, 2018, 2019) also concluded that detonation speed deficit is a function of particle size and concentration. Generally, smaller particle sizes and/or higher particle concentrations tend to reduce the shock speed. These conclusions are also confirmed in Refs. (Fedorov & Tropin 2013; Tropin & Fedorov 2014). Furthermore, some detonation suppression calculations conducted by Tropin and Fedorov (2014) demonstrated that at smaller particle concentrations, particles of 10 and 100 μm (diameter) have limited influences on the detonation velocity deficit, while the influence becomes larger when critical concentrations are reached. They attributed this to the energy exchanges between the gas and particles. Fomin and Chen (2009) proposed that mechanical and heat equilibria are not achieved between the larger particles and shock, and their detonation suppression efficiency is therefore generally lower than that of smaller particles. Although various effects on the detonation propagation in particle curtains were investigated in these studies, the critical condition for detonation inhibition has not been determined, which is of great importance for practical hydrogen safety measures.

Existence of a particle curtain may induce unsteadiness in detonation propagation. For instance, Pinaev et al. (2015) experimentally observed that the propagation speed of $CH_4/2O_2/N_2$ detonation is non-monotonically reduced in a silica sand curtain with three particle size ranges: 90—120, 120—250, and 250—600 μm. They attributed it to the dual role of the sand particles in decelerating the shocked flow and generating hot spots that induce secondary detonations behind the leading shock. More interestingly, Fedorov and Tropin (2011) found that an increased concentration of silica particles, beyond the critical value, will not result in more efficient detonation inhibition. Moreover, a reduction in the particle volume fraction from the critical value to a certain smaller value leads to less detonation speed deficit, compared with the constant limiting concentration.

Two different DW propagation modes in particle curtains modes were observed, i.e., detonation extinction and transmission (Gottiparthi & Menon 2012; Fedorov & Kratova 2013; Kratovaa & Fedorova 2014; Tropin & Fedorov 2018; Tropin & Fedorov 2019; Tropin & Bedarev 2021). In the detonation extinction mode, the reaction front decouples with the leading shock, and their distance is gradually increased. In the latter mode, the reaction zone is still coupled with the SF, but typically with various degrees of speed reduction. In addition, propagation modes of instantaneous extinction followed by detonation re-initiation (Gottiparthi & Menon 2012; Fedorov & Kratova 2013; Fedorov & Tropin 2013; Tropin & Fedorov 2014; Pinaev, Vasilev & Pinaev 2015) and a galloping detonation near the flammability limit (Tropin & Fedorov 2018; Tropin & Fedorov 2019; Tropin & Bedarev 2021) are also observed, which is caused by "explosion in the explosion" (Tropin & Bedarev 2021). The detonation is re-ignited by hot spots due to gas-particles interactions (Pinaev, Vasilev & Pinaev 2015), whilst a galloping mode is a regularly repeated process with a pulsation phenomenon (Tropin & Bedarev 2021). However, the underlying mechanisms behind these transient detonation dynamics have not been clarified, particularly in terms of the interactions between the detonation wave and solid particles.

In this work, detailed simulations with the Eulerian–Lagrangian method and two-way gas-particle coupling are conducted to simulate the transmission of hydrogen detonation in a curtain of dilute inert particles. A two-dimensional configuration is considered and a detailed chemical mechanism is used for hydrogen combustion in our simulations. The objectives of this paper include: (1) critical curtain conditions for quenching detonation; (2) transient detonation phenomena with various particle sizes, concentrations, and curtain thickness; (3) interactions between gas and particles; and (4) evolution of curtain morphology in detonated flows. This paper is structured as follows. The physical model and the mathematical model are introduced in Section 2 and Section 3, respectively. The results and discussion are presented in Section 4, followed by the main conclusions given in Section 5.

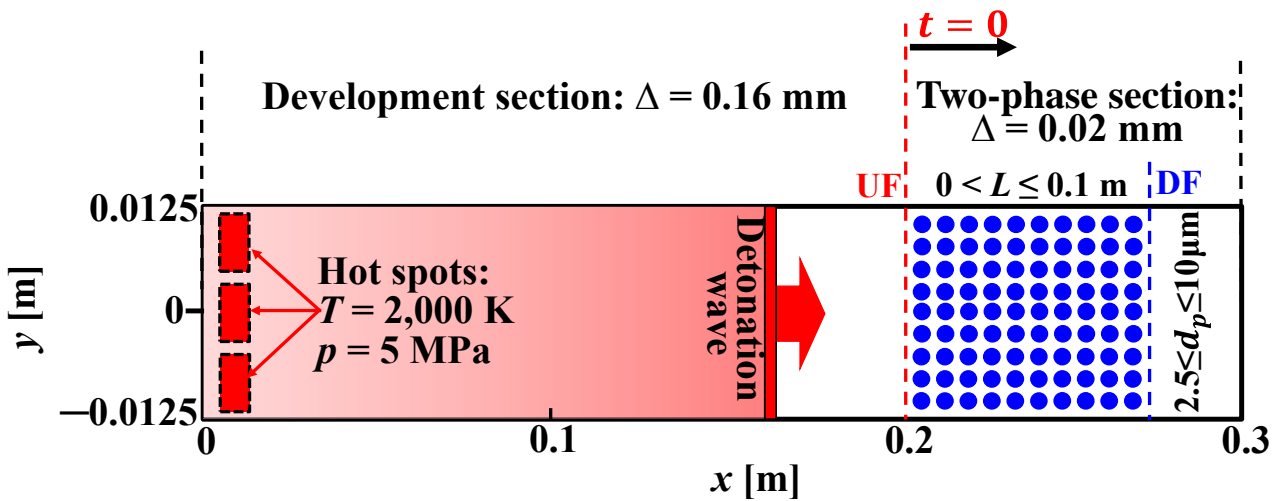

Figure 1 Schematic of the computational domain. Curtain thicknesses $L$ is varied from 0 to 0.1 m. The blue dots represent particles. Domain and particle sizes not to scale. UF and DF: upstream/downstream fronts of the curtain.

## 2. Physical problem

Transmission of a hydrogen detonation wave across a curtain of chemically inert solid particles is simulated with a two-dimensional configuration in Fig. 1. The computational domain is 0.3 m × 0.025 m, which includes a detonation development section (0 ≤ x < 0.2 m) and a two-phase section (0.2 ≤ x ≤ 0.3 m). The domain is initially filled with stoichiometric H$_2$/air mixture. The initial temperature and pressure are $T_0$ = 300 K and $p_0$ = 0.05 MPa, respectively. For the left boundary (x = 0), a non-reflective condition is enforced for the pressure, while a zero-gradient condition for other quantities. Zero-gradient condition is applied at x = 0.3 m and the upper and lower boundaries are set to be periodic.

The two sections are respectively discretized by uniform Cartesian cells of 0.16×0.16 and 0.02×0.02 mm$^2$. The total mesh number is about 8 million. Based on the ZND structure of the particle-free stoichiometric H$_2$/air detonation, the Half-Reaction Length (HRL) $\Delta_{HRL}$ is approximately 382 μm. Therefore, in the two-phase section, at least 19 cells exist within $\Delta_{HRL}$, given that the particle heating behind the lead shock would delay the ignition of shocked gas and hence increase the HRL. Halving the mesh resolution almost does not change the detonation propagation speed and average detonation cell size; see the analysis in Section A of supplementary document.

Monodispersed and static ($\mathbf{u}_p = 0$) solid particles are uniformly distributed in the two-phase section (i.e., x = 0.2 m — 0.2+L m; 0 < L ≤ 0.1 m). The left and right fronts are termed as upstream front (UF) and downstream front (DF), respectively (see the annotation in Fig. 1). In this study, $t = 0$ corresponds to the instant when the detonation reaches the curtain UF. In our simulations, each CFD cell in the curtain area initially has one parcel, and particle concentration variation is achieved by changing the particle number in a parcel. The initial particle concertation of c = 0.1 — 2.5 kg/m$^3$ and particle sizes of $d_p$ = 2.5, 5, and 10 μm will be considered. The size does not change throughout the simulations because of inert particles without any swelling. The two-phase flow is considered to be dilute since the particle concentration does not exceed 0.1% (Crowe et al. 2011). The initial temperature, material density and isobaric heat capacity of the particles are 300 K, 2,500 kg/m$^3$ and

900 J/kg·K, respectively.

## 3. Mathematical model

### 3.1 Gas phase

The governing equations of mass, momentum, energy, and species mass fraction are solved for the multi-component, compressible, reactive flows (Crowe et al. 2011; Zhou et al. 2010):

$$\frac{\partial(\alpha\rho)}{\partial t} + \nabla \cdot [\alpha\rho\mathbf{u}] = 0, \tag{1}$$

$$\frac{\partial(\alpha\rho\mathbf{u})}{\partial t} + \nabla \cdot [\mathbf{u}(\alpha\rho\mathbf{u})] + \alpha\nabla p + \nabla \cdot (\alpha\mathbf{T}) = \mathbf{S}_{\text{mom}}, \tag{2}$$

$$\frac{\partial(\alpha\rho E)}{\partial t} + \nabla \cdot [\mathbf{u}(\alpha\rho E + \alpha\mathbf{p})] + \nabla \cdot [\alpha\mathbf{T} \cdot \mathbf{u}] + \nabla \cdot (\alpha\mathbf{j}) + p\frac{\partial \alpha}{\partial t} = \alpha\dot{\omega}_T + S_{energy}, \tag{3}$$

$$\frac{\partial(\alpha\rho Y_m)}{\partial t} + \nabla \cdot [\mathbf{u}(\alpha\rho Y_m)] + \nabla \cdot (\alpha\mathbf{s_m}) = \alpha\dot{\omega}_m \; (m = 1, \ldots, M-1). \tag{4}$$

Here $t$ is time and $\nabla \cdot (\cdot)$ is the divergence operator. $\alpha$ is the gas phase volume fraction, $\rho$ is the gas density, $\mathbf{u}$ is the gas velocity vector, and $T$ is the gas temperature. $p$ is the pressure, derived from the ideal gas equation of state. $\mathbf{T}$ is the viscous stress tensor. In Eq. (3), $E$ is the total non-chemical energy, and $\mathbf{j}$ is the diffusive heat flux. The term $\dot{\omega}_T$ represents the heat release rate from chemical reactions. In Eq. (4), $Y_m$ is the mass fraction of $m$-th species, $M$ is the total species number, and $\mathbf{s_m}$ is the species mass flux. $\dot{\omega}_m$ is the production or consumption rate of $m$-th species by all reactions. The source terms in Eqs. (2) and (3), i.e., $\mathbf{S}_{mom}$, and $S_{energy}$, denote the exchanges of momentum and energy between gas and particles. Their corresponding expressions are given in Eqs. (13) and (14), respectively. Details of the gas phase equations can be found in Refs. (Xu, Zhao & Zhang 2021; Xu & Zhang 2022).

### 3.2 Particulate phase

The Lagrangian method is used to track a number of spherical solid particles in the dispersed phase. The particle collisions are neglected because of dilute particle concentrations (Crowe et al. 2011). Since the ratio of gas density to the particle material density is well below one, the Basset force, history force, and gravity force are not considered (Crowe et al. 2011). Therefore, the equations of momentum and energy of a single particle read

$$\frac{d\mathbf{u}_p}{dt} = \frac{\mathbf{F}_{p,d} + \mathbf{F}_{p,p}}{m_p}, \qquad (5)$$

$$c_{p,p}\frac{dT_p}{dt} = \frac{\dot{Q}_c}{m_p}, \qquad (6)$$

where $m_p = \pi \rho_p d_p^3/6$ is the mass of a single particle, which is constant. $\rho_p$ and $d_p$ are the particle material density and diameter, respectively. $\mathbf{u}_p$ is the particle velocity vector, $c_{p,p}$ is the particle heat capacity, and $T_p$ is the particle temperature.

The drag force $\mathbf{F}_{p,d}$ in Eq. (5) is modelled as (Liu, Mather & Reitz 1993)

$$\mathbf{F}_{d,p} = \frac{18\mu}{\rho_p d_p^2} \frac{C_d Re_p}{24} m_d (\mathbf{u} - \mathbf{u}_p). \qquad (7)$$

The drag coefficient, $C_d$, is estimated with (Liu, Mather & Reitz 1993)

$$C_d = \begin{cases} 0.424, & \text{if } Re_p \geq 1000, \\ \frac{24}{Re_p}\left(1 + \frac{1}{6}Re_p^{2/3}\right), & \text{if } Re_p < 1000. \end{cases} \qquad (8)$$

The particle Reynolds number in Eq.(8), $Re_p$, is defined as

$$Re_p \equiv \frac{\rho d_p |\mathbf{u}_p - \mathbf{u}|}{\mu}. \qquad (9)$$

Here $\mu$ is the gas dynamic viscosity. Besides, the Archimedes force (or pressure gradient force) $\mathbf{F}_{p,p}$ in Eq. (5) takes the following form

$$\mathbf{F}_{p,p} = -V_p \nabla p, \qquad (10)$$

where $V_p$ is the volume of a single particle.

The convective heat transfer rate $\dot{Q}_c$ in Eq. (6) is

$$\dot{Q}_c = h_c A_p (T - T_p). \qquad (11)$$

Here $A_p$ is the particle surface area, and $h_c$ is the convective heat transfer coefficient, following

Ranz and Marshall (Ranz & Marshall 1952)

$$Nu = h_c \frac{d_p}{k} = 2.0 + 0.6 Re_p^{1/2} Pr^{1/3}, \tag{12}$$

where $Nu$ and $Pr$ are the Nusselt and Prandtl numbers of the gas phase, respectively.

The particle effects on the gas phase are modelled with Particle-source-in-cell approach (Crowe, Sharma & Stock 1977), which are modelled with the source terms, i.e., $\mathbf{S}_{mom}$ and $S_{energy}$ for Eqs. (2) and (3). They read

$$\mathbf{S}_{mom} = -\frac{1}{V_c}\sum_{i=1}^{N_p}(\mathbf{F}_{p,d,i} + \mathbf{F}_{p,p,i}), \tag{13}$$

$$S_{energy} = -\frac{1}{V_c}\sum_{i=1}^{N_p}\dot{Q}_{c,i}. \tag{14}$$

$V_c$ is the CFD cell volume and $N_p$ is the particle number in one CFD cell.

### 3.3 Computational method

The equations for gas particle phases are solved by a two-phase compressible reactive flow solver, *RYrhoCentralFoam* (Huang et al. 2021; Zhao & Zhang 2020), based on OpenFOAM 6.0 (Weller et al. 1998). The simulations run with 3,840 processors of the *Fugaku* Cluster from the RIKEN Center for Computational Science in Japan. The physical time of 100 microsecond can be achieved with wall clock time of approximately 22 hours.

For gas phase equations, second-order backward method is employed for temporal discretization and the time step is about $2 \times 10^{-9}$ s. A Riemann-solver-free MUSCL scheme (Kurganov, Noelle & Petrova 2001) with van Leer limiter is used for convective fluxes in the momentum equations, and total variation diminishing scheme is used for the convection terms of energy and species equations. Second-order central differencing scheme is applied for the diffusion terms in Eqs. (2)−(4). The detailed mechanism (13 species and 27 reactions) by Burke et al. (2012) is used for hydrogen combustion.

For the Lagrangian phase, the particles are tracked based on their barycentric coordinates. Equations (5) and (6) are integrated by first-order implicit Euler method. Meanwhile, the gas properties (e.g., velocity and temperature) at the particle location are calculated based on linear interpolation.

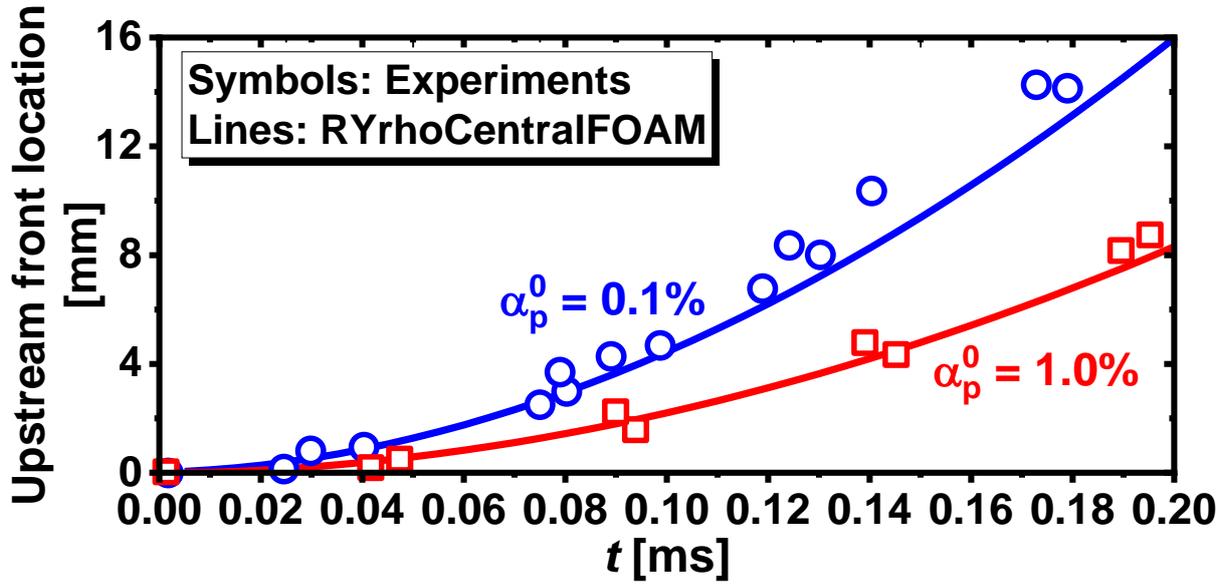

Figure 2 Time history of the upstream front location after the shock crosses the bronze particle curtain. The initial particle volume fractions: blue: 0.1%; red: 1.0%. Experimental data: Boiko et al. (1997).

**3.4 Solver validation**

The *RYrhoCentralFoam* solver has been validated and verified for both gaseous and gas-droplet two-phase detonations (Huang et al. 2021; Zhao & Zhang 2020), and have been successfully applied for a range of detonation and supersonic combustion problems (Xu & Zhang 2022; Zhao & Zhang 2020; Zhao, Ren & Zhang 2021; Zhao et al. 2020; Zhao & Zhang 2020; Zhao & Zhang 2021; Meng et al. 2020; Zhao, Chen & Zhang 2021; Zhu, Zhao & Zhang 2021). Here we further validate the solver for shock-curtain interaction experiments by Boiko et al. (1996), in which the configuration is similar to the one of our studies. One-dimensional domain is considered, filled with helium. The incident shock Mach number is 2.8 (corresponding shock speed is 2,854 m/s). The initial pressure and temperature of the driven gas $p_0 = 0.1$ MPa and $T_0 = 300$ K, respectively. The curtain is composed of spherical bronze particles. The material density and heat capacity are $8,600 \text{ kg/m}^3$ and $435 \text{ J/(kg·K)}$, respectively. The initial diameter is $d_p = 130 \text{ μ}m$. Two initial volume fractions are considered, i.e., 0.1% and 1.0%. Uniform cells of 500 μm are used, and particles are evenly distributed across the 13 mm curtain. As shown in Fig. 2, our solver can accurately reproduce the trajectories of curtain UF after shock transmission in both cases. Therefore, the solver accuracies in predicting shock-curtain interaction can be confirmed.

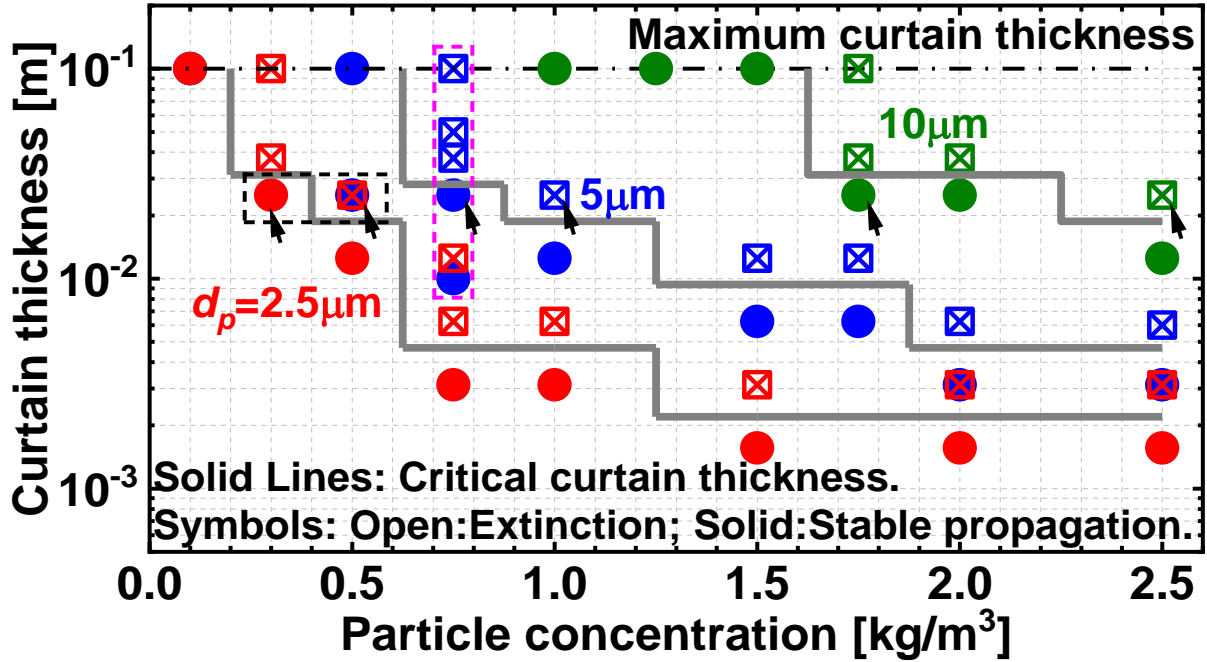

Figure 3 Diagram of detonation propagation and extinction across an inert particle curtain. The solid curves denote the critical curtain thickness $L_{cr}$. The dash-dotted line is the length of the two-phase section (0.1 m).

## 4. Results and discussion

### 4.1 Detonation regime map

Unsteady behaviors of detonation propagation in a particle curtain are evaluated by parametric studies, and the results are summarized in Fig. 3, through the diagram of initial particle concentration $c$ versus curtain thickness $L$. Three particle diameters are considered, i.e., 2.5, 5, and 10 μm. The critical curtain thicknesses $L_{cr}$, denoted by the solid lines, correspond to the critical cases between a transmission and extinction case, which are determined based on a trial-and-error method. The reader should be reminded that the y-axis is logarithmic, and hence the curve seems step-wise.

In general, $L_{cr}$ decreases quickly with the particle concentration $c$ for all three diameters. For instance, if we look at $d_p = 2.5$ μm, when $c$ is 0.5 kg/m³, the critical length is about 20 mm, whereas when $c =$ 2.0, it is reduced to only approximately 2.2 mm. This is because higher particle concentration corresponds to larger transfer of momentum and energy, and a shorter curtain thickness is required for quenching a propagating detonation wave. Beyond some critical particle concentration (e.g., > 1.3 kg/m³), $L_{cr}$ exhibits weak dependence on $c$, which may be limited by the finitely long timescales of momentum and energy exchange between two phases. This tendency is also revealed from methane detonation inhibition with ultrafine water spray curtain by Shi et al. (2022).

Moreover, with increased particle size, the concentration range to have detonation extinction is appreciably elevated when the curtain thickness is fixed (e.g., 0.01 m). This is because smaller particles with larger specific surface area are more efficient to weaken the detonation wave through faster heat and momentum absorption. Furthermore, with the same concentration, e.g., 2 kg/m³, the critical curtain thickness $L_{cr}$ decreases when $d_p$ becomes smaller. For instance, when $d_p =$ 2.5 and 10 μm, their corresponding $L_{cr} =$ 2.2 and 31 mm with $c =$ 2 kg/m³.

The implications of the results in Fig. 3 for practical implementation of inert dust for hydrogen detonation inhibition are multi-fold. Firstly, smaller particles are better choice because of their wider concentration range to quench a detonation. Secondly, when the curtain thicknesses are the same, the particle concentration required to quench a detonation increases significantly with the particle diameter. Thirdly, when the particle size is fixed, short curtain with high particle concentration would also be a solution, but further increasing the concentration would not further shorten the curtain dimension.

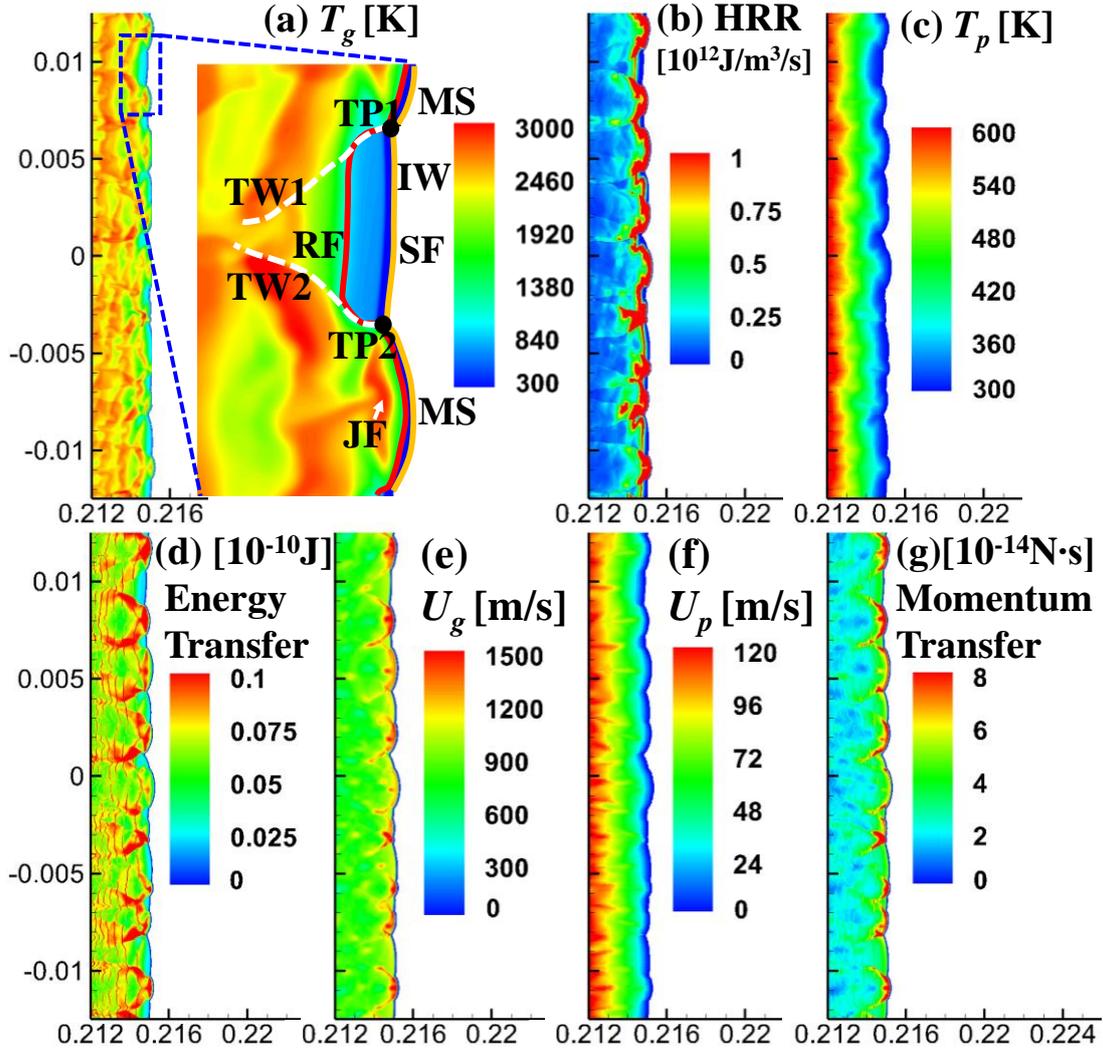

Figure 4 Contours of (a) gas temperature, (b) heat release rate, (c) particle temperature, (d) energy exchange, (e) gas velocity, (f) particle speed, and (g) momentum transfer. HRR: heat release rate. RF/SF: reaction/shock front. MS: Mach stem. IW: incident wave; TW: transverse wave. TP: triple point. JF: jet flow. Axial label unit: m.

### 4.2 Two-phase detonation dynamics

#### 4.2.1 Detonation structure

The instantaneous structure of two-phase detonation in the particle curtain is demonstrated in Fig. 4. The results in Fig. 5 are averaged from those in Fig. 4 along the domain width (*y*-direction) in the curtain. The curtain conditions are $c = 0.75$ kg/m³, $d_p = 5$ μm, and $L = 25$ mm. Note that a negative transfer value in Figs. 4 and 5 indicates that momentum or heat is transferred from the gas to the particles. From the gas temperature contour in Fig. 4(a), the detonation frontal structures, e.g., SF (IW or MS), RF, TW and JF (see their full names in the caption of Fig. 4), are well captured. Moreover, in the induction zone (see enlarged view of Fig. 4a), the gas temperature behind the MS is much higher than that of the IW, due to higher shock wave intensity. Stronger RF, featured by higher heat release rate (HRR) as seen from Fig. 4(b), occurs immediately behind the MS and partial TW (transverse detonation wave), whilst weaker RF is seen behind the IW. One can see from the averaged profiles in Fig. 5(a) that the HRR peaks around the RF, roughly $3\times10^{12}$ $J/m^3/s$. Behind the MS, the gas pressure (not shown here) is higher than that of the IW, and the JF can ignite the shocked gas immediately behind the MS (Bhattacharjee et al. 2013).

Different from the gas temperature distributions of Fig. 4(a), the particle temperature exhibits clear layered distributions, as illustrated in Fig. 4(c). This can be more clearly seen from Fig. 5(b) that both gas and particle temperatures behind the SF increase, but the latter is still lower than the former within the shown distance. Thermal equilibrium is not achieved between two phases in the curtain and therefore interphase energy transfer would occur in Fig. 4(d), as indicated in Eq. (14). In Fig. 4(d), larger heat transfer behind the Mach stem and TW are caused by greater temperature difference in these locations.

Figures 4(e) and 4(f) show that larger velocity difference can be found between the gas and particles. Higher momentum exchange (see Eq. 13) generally appears behind the MS and TW, as shown in Fig. 4(g). This is due to larger gas velocity and larger pressure gradient (see Fig. 8) from the same locations. This velocity difference exists for a long distance after the leading SF, as found from the average profile in Fig. 5(c). Figure 5(c) also shows that their velocities are not equilibrated in the particle curtain. Therefore, the momentum exchange proceeds in the entire curtain and its magnitude peaks inside the induction zone, slightly earlier than that of the energy transfer. This is because the momentum transfer timescale is much shorter than that of the heat exchange. Meanwhile, in the shock-frame, the induction zone transits from supersonic to subsonic zone due to higher gas speeds behind the SF. Behind the RF, the shock-frame Mach number approaches unity, featuring a freely propagating detonation wave, as observed from Fig. 5(a).

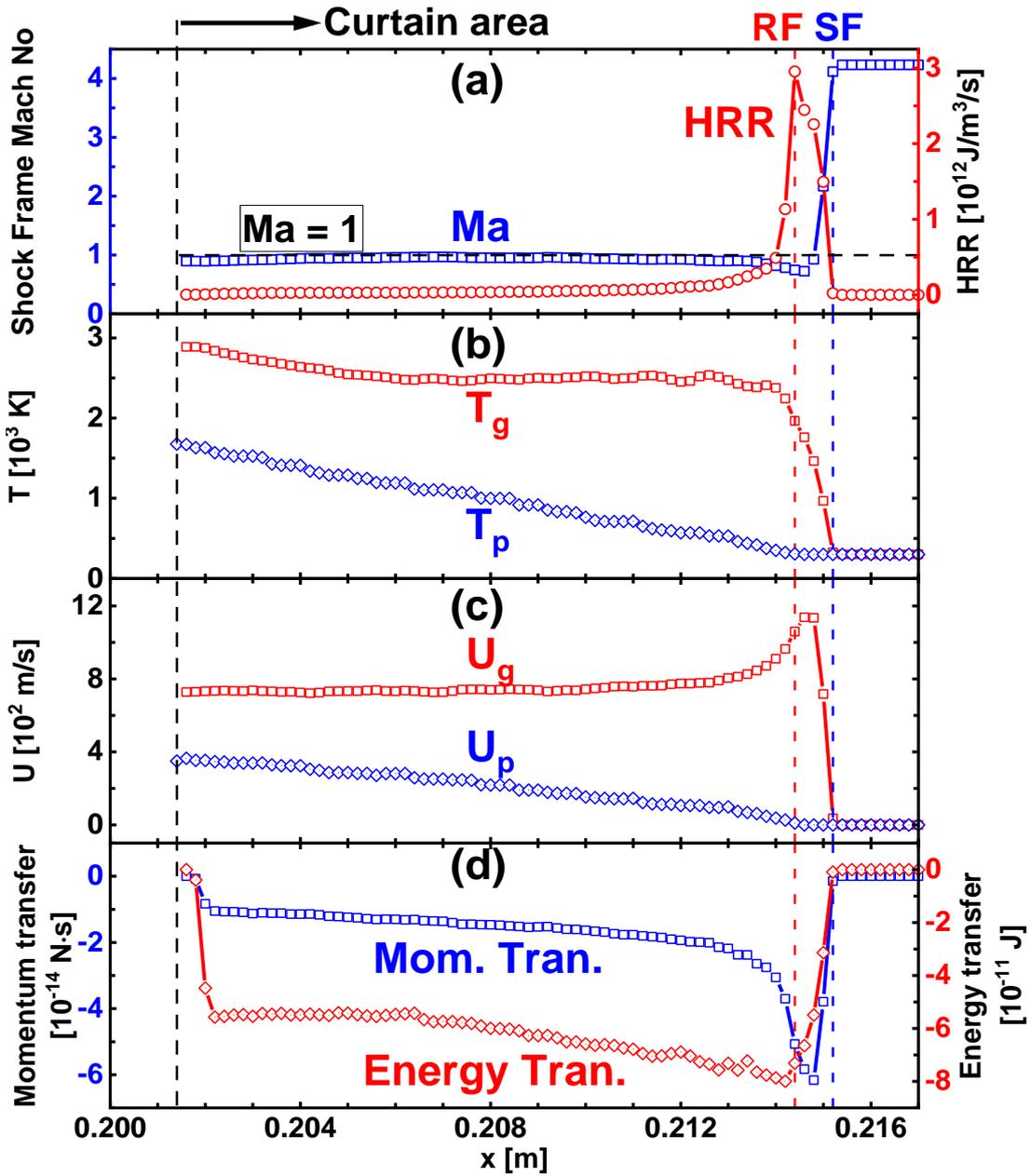

Figure 5 Averaged profiles of key quantities along the detonation front in Fig. 4. RF/SF: reaction/shock front. Arrow: curtain area.

**4.2.2 Unsteady detonation phenomena**

When the particle size or concentration is varied, unsteady detonation phenomena are observed. Plotted in Fig. 6 is a detonation extinction and re-initiation process, demonstrated by the evolutions of gas temperature. The curtain conditions are $c = 0.75$ kg/m$^3$, $d_p = 5$ μm, and $L = 25$ mm. As annotated in Fig. 6, at 3 -13 μs, the DW runs in the gas-particle mixture (inside the curtain). Specifically, at 3 μs, the relatively stable DW just enters the curtain. Subsequently, from 9 to 23 μs, along the detonation front, more sections of the RF are decoupled from the SF, featured by increased distance between them. This is because of the heat and/or momentum absorption by the particles in the post-detonation flow. Although the DW leaves the curtain at around 13 μs, the extent of localized extinction of the detonation wave still increases (e.g., at 18 and 22 μs), manifested by the continuously lengthened the distance between the RF and SF. Nonetheless, there are still some sections of the leading shock front with detonative combustion (e.g., directed by the arrows at 18 and 22 μs). At 22 μs, a re-initiation (RI in Fig. 6) point appears near the upper boundary. The RI transient will be further discussed in Figs. 7 and 8. The RI reaction front further evolves into an overdriven detonation wave from 23 μs. It quickly develops into longitudinally and transversely propagating waves. The latter consumes the un-reacted gas between the decoupled RF and SF. After 41 μs, the detonation front has restored to a stable propagation state.

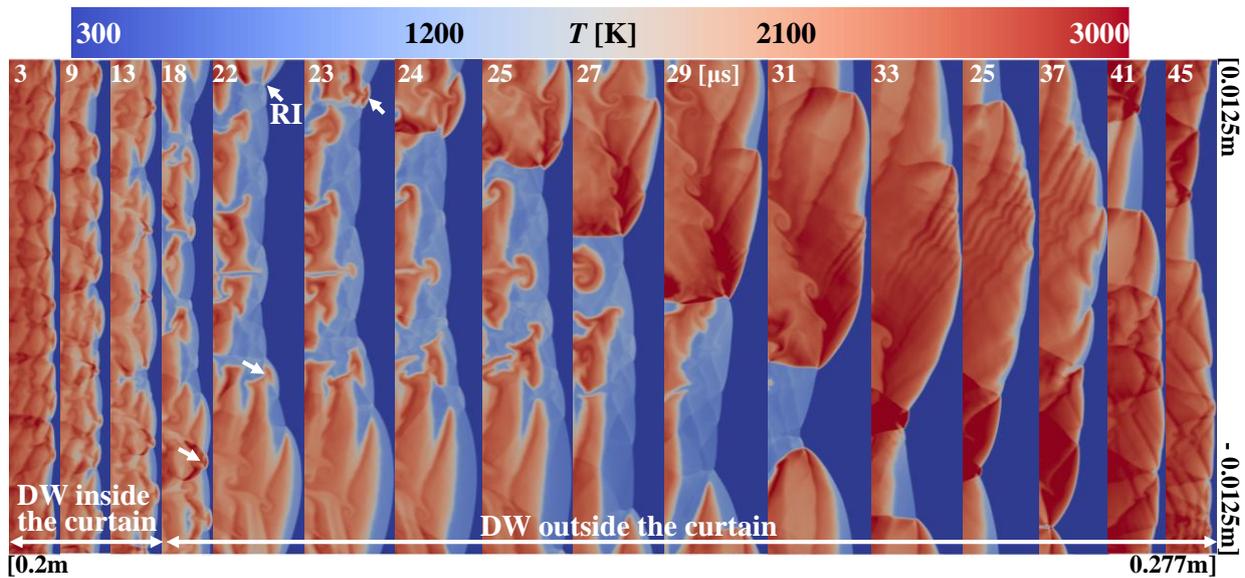

Figure 6 Time history of shock frontal structures in a detonation re-initiation process. $c = 0.75$ kg/m$^3$, $d_p = 5$ μm, and $L = 25$ mm. RI: re-initiation.

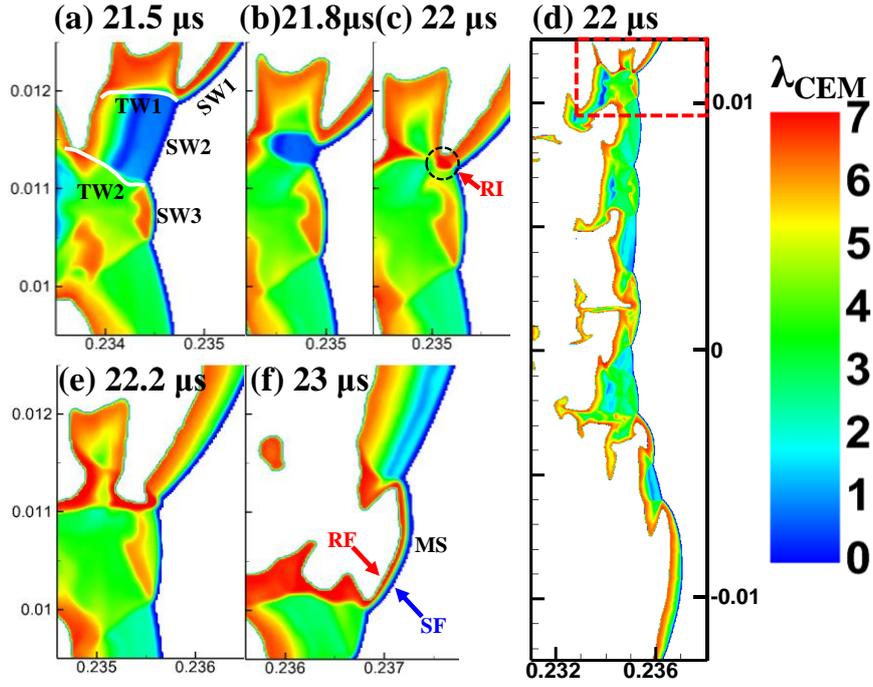

Figure 7 Distributions of $\lambda_{CEM}$ in a detonation re-initiation process. $c$ = 0.75 kg/m$^3$, $d_p$ = 5 μm, and $L$ = 25 mm. TW: transverse wave. SW: shock wave. RI: Re-initiation. Axial label unit: m.

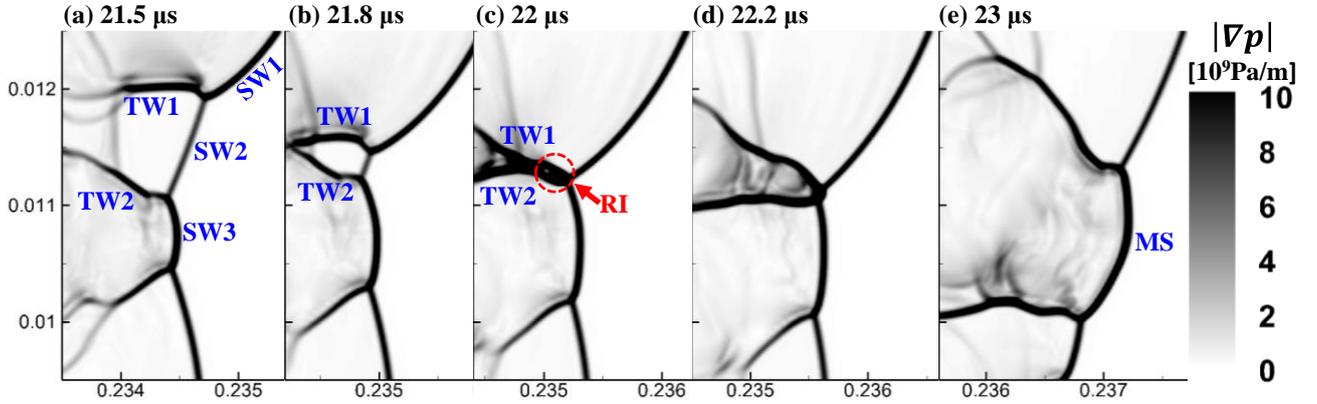

Figure 8 Distributions of pressure gradient magnitude in a partial extinction and detonation re-initiation process corresponds to Fig. 7. $c$ = 0.75 kg/m$^3$, $d_p$ = 5 μm, and $L$ = 25 mm. TW: transverse wave. SW: shock wave. RI: Re-initiation point. Axial label unit: m.

To quantify the chemical structure evolutions of a detonation partial extinction and re-initiation process, we calculate the chemical explosive mode (Lu et al. 2010; Wu et al. 2019; Lam 1985; Lam & Goussis 1998; Lam 1993; Lam 2007) of the detonable gas between the SF and RF for the results in Fig. 6. Figure 7 shows the exponent of the eigenvalue $\lambda_e$ of chemical explosive mode, i.e., $\lambda_{CEM} = sign[Re(\lambda_e)] \cdot log_{10}[1 + |Re(\lambda_e)|]$. It is associated with the chemical timescale and directly reflects the local chemical reactivity of gaseous mixture. Enlarged view of Figs. 7 (a—c and e—f) are extracted from the dashed box in Fig. 7(d). Note that positive $\lambda_{CEM}$ corresponding to the propensity of chemical explosion (Lu et al. 2010). For clear illustration, only positive $\lambda_{CEM}$ is shown in Fig. 7. Figure 8 shows the distributions of pressure gradient magnitude around the RI point in Fig. 6.

At 21.5 μs, highly / weakly explosive gas mixture is found behind the SW1 and SW3 / SW2, featured by higher / smaller $\lambda_{CEM}$ ( above 4 / below 2). Meanwhile, a certain distance between TW1 and TW2 can be seen in Fig. 7(a) or 8(a). From 21.5 to 21.8 μs, the two TWs propagate towards each other and intersect at 22 μs in Fig. 8(c). This transverse shock focusing leads to the formation of a RI point with high pressure and $\lambda_{CEM}$ in Fig. 7(c). Since 22.2 μs, a new Mach stem is developed from this RI point, demonstrated with the coupled RF and SF in Fig. 7(f). It is also evidenced by larger $\lambda_{CEM}$ before the RF.

Figure 9 shows the time evolutions of shock frontal structures during a detonation extinction event from Fig. 3. It has the same particle conditions to Figs. 6—8, but a longer curtain thickness, i.e., $L = 37.5$ mm. Note that the DW is inside the curtain till 23 us, i.e., all the instants in Fig. 9 correspond to gas-particulate medium. Specifically, at 3 μs, the multi-headed DW is observed when it just encroaches the 37.5 mm curtain. Subsequently, from 8 to 20 μs, partial decoupling of the DW behind the leading SF can be seen, and the decoupled portion becomes greater. This process is accompanied by the increased distance between the SF and RF at most locations of the leading front, and the reduced gas temperature behind the leading SF. The SF and RF are fully decoupled after 22 μs.

Likewise, the evolutions of the corresponding chemical explosive mode in this detonation extinction event are shown in Fig. 10. The instant in Fig. 10(a) presents that highly explosive mixture still occupies the induction zone, demonstrated with $\lambda_{CEM} > 5$. Subsequently, due to the decoupling process in Fig. 10 (b), $\lambda_{CEM}$ is significantly reduced to below 5 behind the SF, whilst high $\lambda_{CEM}$ is only observed immediately ahead of the RF. This demonstrates that the explosion propensity in the induction zone is reduced. Finally, the RF is completely decoupled from the SF, and $\lambda_{CEM}$ in most portion of the RF—SF structures is reduced to below 2. Therefore, the weakly explosive mixture dominates between the RF and SF.

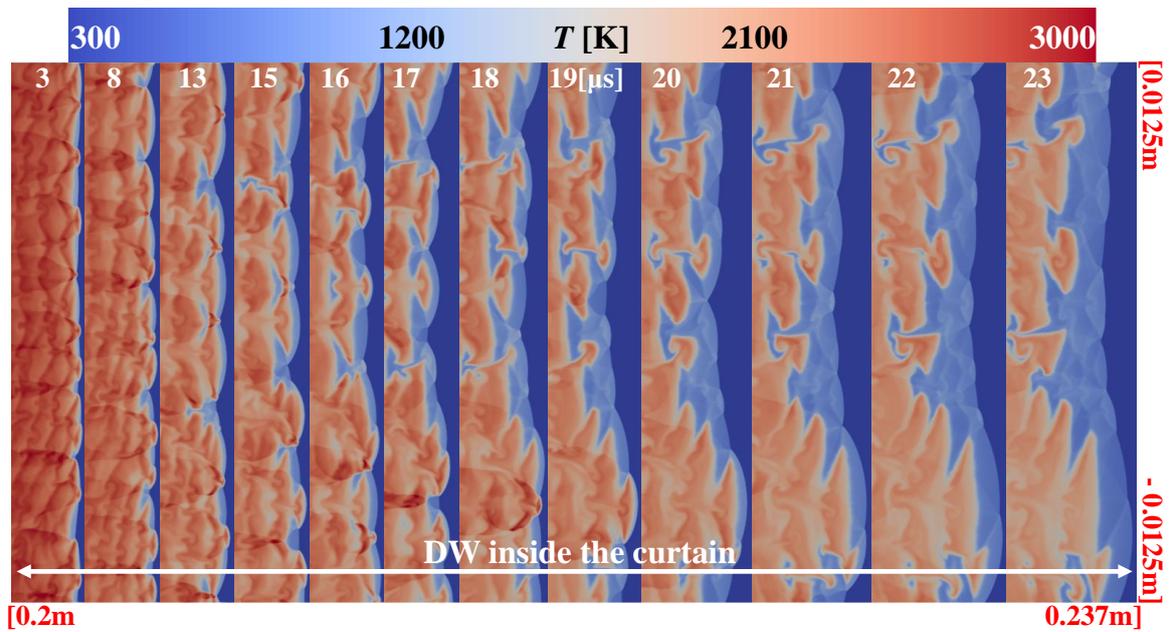

Figure 9 Time history of shock frontal structures in a detonation extinction process. $c = 0.75$ kg/m$^3$, $d_p = 5$ μm, and $L = 37.5$ mm.

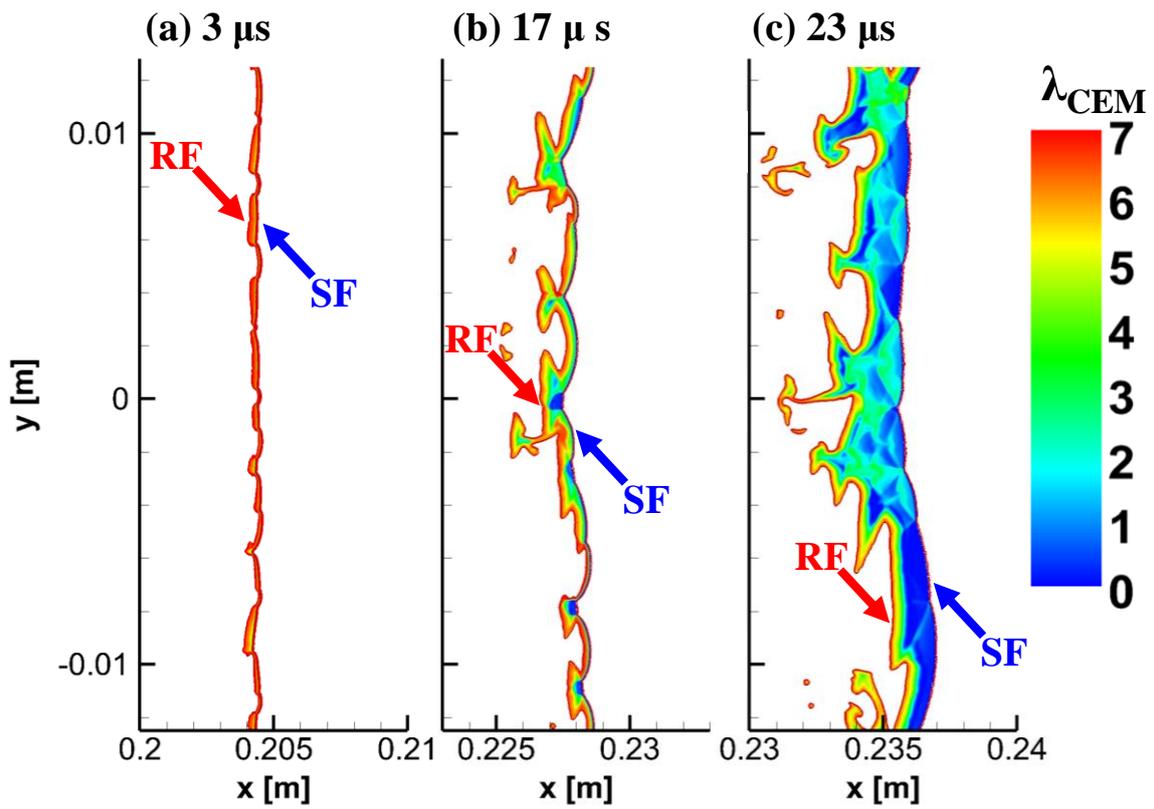

Figure 10 Distributions of $\lambda_{CEM}$ in an immediate detonation extinction process. $c = 0.75$ kg/m$^3$, $d_p = 5$ μm, and $L = 37.5$ mm. Left-/Rightmost lines: Shock/Reaction Front (SF/RF).

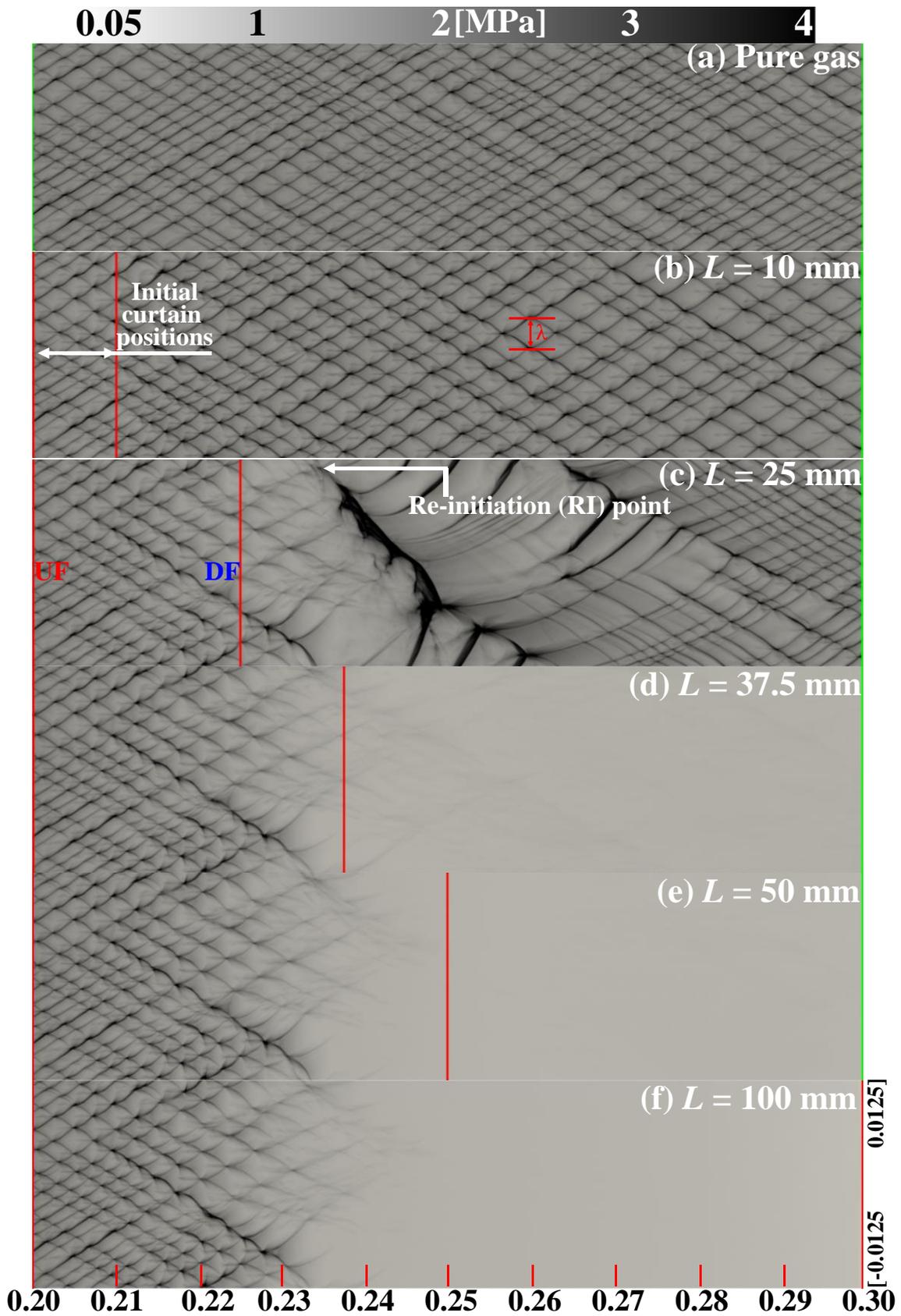

Figure 11 Peak pressure trajectories with various particle curtain thicknesses. The red lines denote the initial curtain positions. UF/DF: upstream/downstream front. $c$ = 0.75 kg/m$^3$ and $d_p$ = 5 μm. Axis label unit: m.

### 4.2.3 Effects of curtain thickness

In this section, the curtain properties (i.e., curtain thickness, particle diameter and concentration) on the hydrogen detonation wave will be examined. Figure 11 shows the peak pressure trajectories with six curtain thicknesses, i.e., 10 − 100 mm. They correspond to the cases in the pink box in Fig. 3, and the corresponding particle concentration and diameter are $c$ = 0.75 kg/m³ and $d_p$ = 5 μm, respectively. For comparison, the curtain-free case is also included in Fig. 11(a). The cellular structures of the detonation wave are significantly changed when it is transmitted in the curtain. Note that the mismatch of the acoustic impedance (i.e., product of density and sound speed) across the interface (i.e., UF or DF) between the gaseous mixture and curtain is small (approximately 256 kg/m²s) (Ranjan, Oakley & Bonazza 2011). Therefore, reflected shocks or expansion waves from the UF and DF of the curtain are not observable from our results.

For the shortest curtain in Fig. 11(b), the averaged cell width behind the curtain (2.9 mm) is slightly increased (e.g., at $x$ = 0.26 − 0.3 m) compared to that in Fig. 11(a) (1.7 mm). This indicates the increased detonation instability in the post-curtain area. We term this as detonation transmission mode.

Furthermore, with $L$ = 25 mm in Fig. 11(c), after the DW propagates a finitely long distance inside the curtain, the peak pressure trajectories gradually fade at $x$ = 0.22 − 0.26 m. In fact, the continuous propagation of the DW is decoupled into separated SF and RF after $x$ = 0.22 m. Nonetheless, a re-initiation (RI) point appears at around $x$ = 0.236 m, and a thick trajectory evolves from there. Clear cellular structures are restored after $x$ = 0.28 m. We term this as partial extinction and detonation re-initiation mode.

When the curtain thickness is further increased beyond 25 mm, as in Figs. 11(d)-11(g), the incident DWs are quenched inside the curtain, characterized by quickly reduced peak pressures left to the DF. Different from Fig. 11(c), no RI events are observed. It corresponds to the third mode, i.e., detonation extinction. Note that the extinction distance in these cases (37.5 − 100 mm) are close, e.g., around 32 mm relative to the curtain UF.

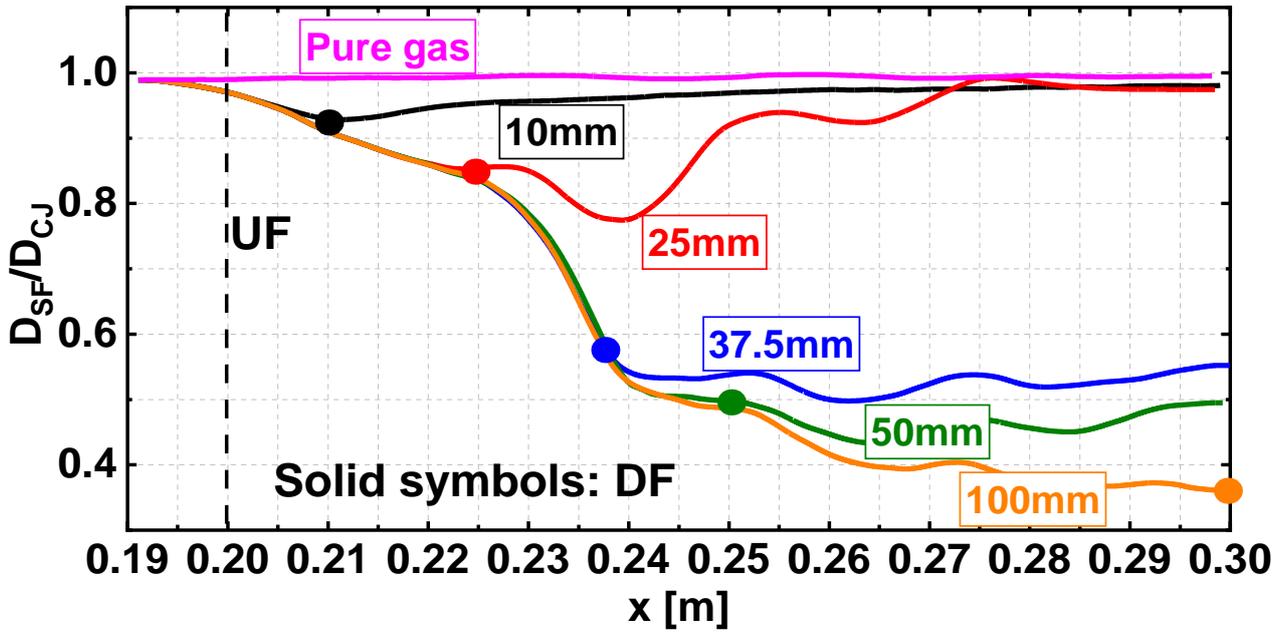

Figure 12 Effects of curtain thickness on shock velocities. $D_{CJ}$ = 1961 m/s. $c$ = 0.75 kg/m³, and $d_p$ = 5 μm. The dashed line and dots respectively denote the initial locations of curtain UF and DF before detonation impacting.

Figure 12 further quantifies the evolutions of lead shock speed $D_{SF}$ corresponding to the cases in Fig. 11. They are scaled by the C—J speed $D_{CJ}$ (1,961 m/s) of particle-free hydrogen/air mixture. A slightly reduced shock speed is observed with the 10 mm curtain, and after the shock leaves the curtain, it is gradually restored to approximately the C—J value. When the DW crosses the longer curtain of $L$ = 25 mm, the velocity drops to below 80% $D_{CJ}$ around $x$ = 0.24 m. Thereafter, due to a detonation re-initiation event discussed above, $D_{SF}$ rises to around the C—J value at around $x$ = 0.275 m. Nonetheless, when $L$ > 37 mm, all the shock speeds are monotonically reduced to below $0.6D_{CJ}$ before $x$ = 0.24 m. However, after $x$ = 0.24 m, their speeds level off between $0.3-0.6$ $D_{CJ}$ when the SF/RF are fully decoupled.

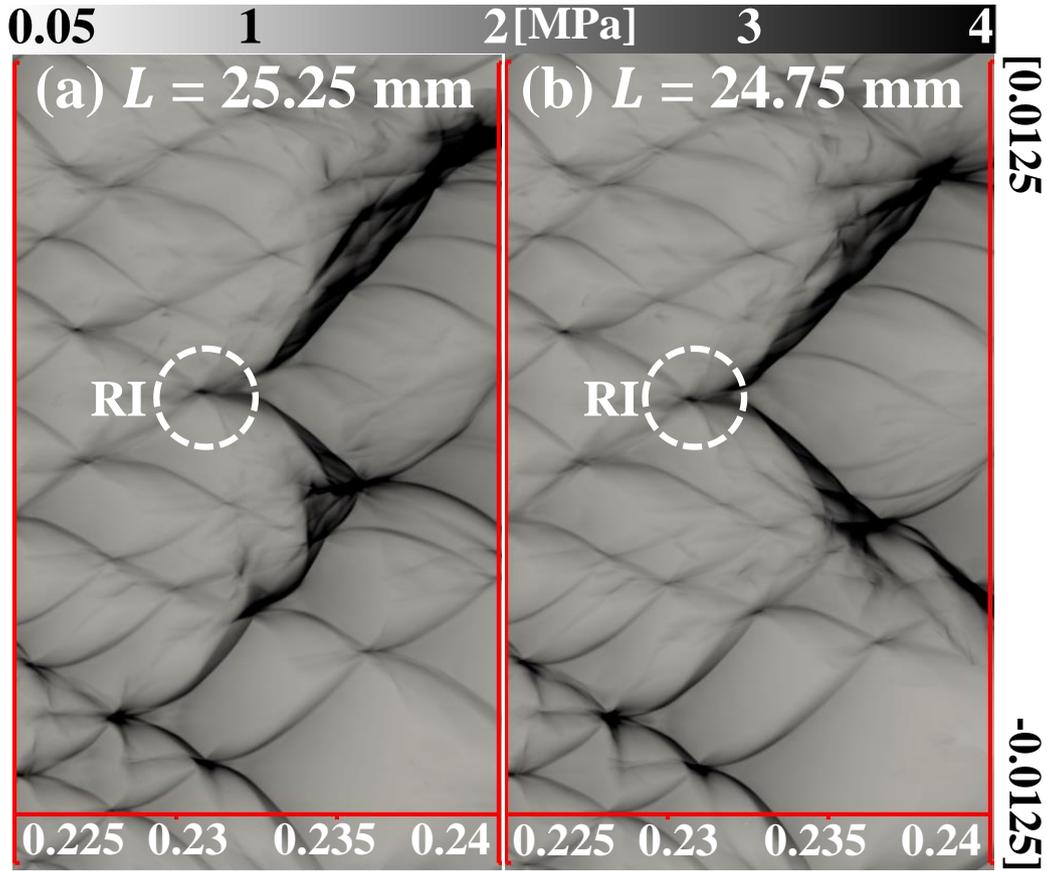

Figure 13 Trajectories of peak pressure with perturbed curtain thicknesses: (a) 25.25 mm and (b) 24.75 mm. $c$ = 0.75 kg/m$^3$, and $d_p$ = 5 μm. RI: re-initiation point. Axis label unit: m.

To further examine the stochasticity of occurrence and location of the RI event in Fig. 11(c), we conduct two numerical experiments through perturbing the curtain thickness in that case (25 mm) by ±1%, i.e., 25.25 and 24.75 mm. Their corresponding trajectories are plotted in Fig. 13. It is observed that RI phenomenon still occurs when the curtain thickness is slightly perturbed. Nonetheless, different from the RI location in Fig. 11(c), here both RI points (labelled with the circles) occur around the domain centerline. This implies that the RI occurrence is not a fully stochastic event, but the RI location is very sensitive to the curtain dimension. Moreover, small length perturbation (±1%) has significant influences on the time history of HRR, as shown in Fig. 14. Note that HRR is volume-averaged based on the two-phase section of $x$ = 0.2―0.3 m. From 19 to 43 μs, corresponding to perturbed regime, their HRR profiles from the perturbed curtains largely deviate from the results of 25 mm curtain. However, beyond 43 μs, the HRR profiles are almost the same in the restoration period. Note that the SF at 19 and 43 μs are respectively located at $x$ = 0.231 m and 0.273 m. It is also observed from the restoration period that the average detonation cell size and cellularity (not given here) are almost the same. Therefore, small variations of the curtain thickness would affect the short-term behaviors of detonation development, but the long-term DW behaviors are negligibly affected. The implications of Figs. 13 and 14 are that small curtain dimension difference would lead to possible change of critical detonation phenomena, which therefore poses higher accuracy requirement for predictability of detonation and particle interactions.

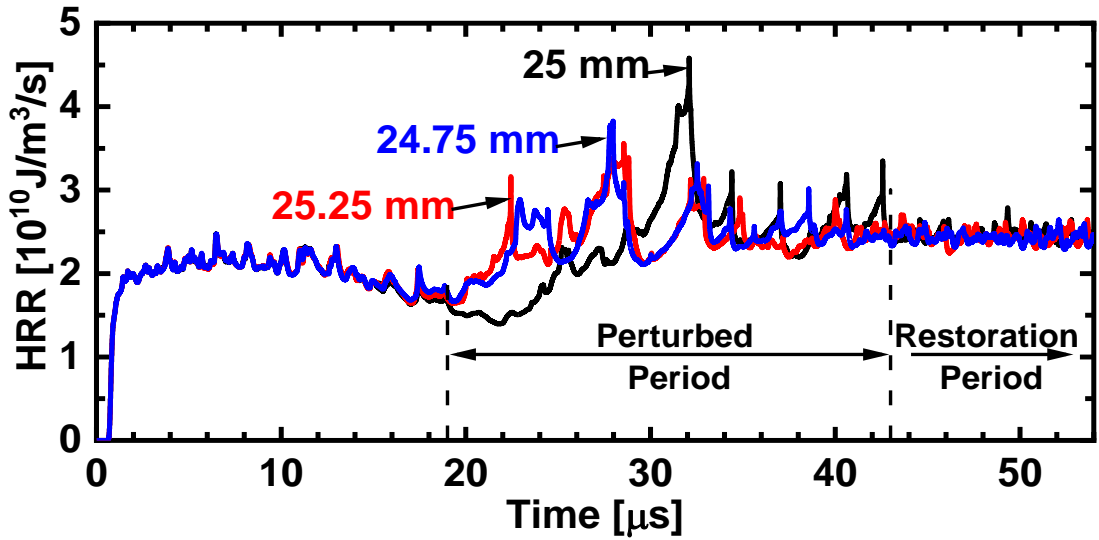

Figure 14 Time history of heat release rate with perturbed curtain thickness. $c$ = 0.75 kg/m³ and $d_p$ = 5 μm.

### 4.2.4 Effects of particle diameter and concentration

The influences of particle diameter and concentration on peak pressure trajectories of hydrogen detonation are shown in Fig. 15. These cases are from the black box in Fig. 3, and have the same curtain thickness, i.e., $L$ = 25 mm. Particle concentration effects are examined through Figs. 15(a) and 15(b) since they have the same diameter, whilst Figs. 15(b) and 15(c) are used to study the droplet diameter effects. Additionally, the evolutions of the shock speeds in the foregoing cases are plotted in Fig. 16. It is found from Fig. 15(a) that the low-concentration case only exhibits localized extinction and a secondary DW re-initiation. The triple point trajectories are also recovered after a streamwise distance of 44 mm, almost the same as Fig. 11(c). This is mainly because the shock re-initiates in the same pure gas environment. Meanwhile, following this curtain condition, the shock speed in Fig.16 (red line) drops to below $0.85D_{CJ}$ but quickly increases to $0.98D_{CJ}$ due to a re-initiation event. However, in Fig. 15(b), large concentration of the particles can quench the DW within the curtain, featured by fading trajectories of the triple points in the two-phase section. This detonation quenching process is also characterized by the significantly decreased velocity (blue line in Fig. 16): as low as $0.55D_{CJ}$ at $x$ = 0.225 m and thereafter levels off at $0.6D_{CJ}$.

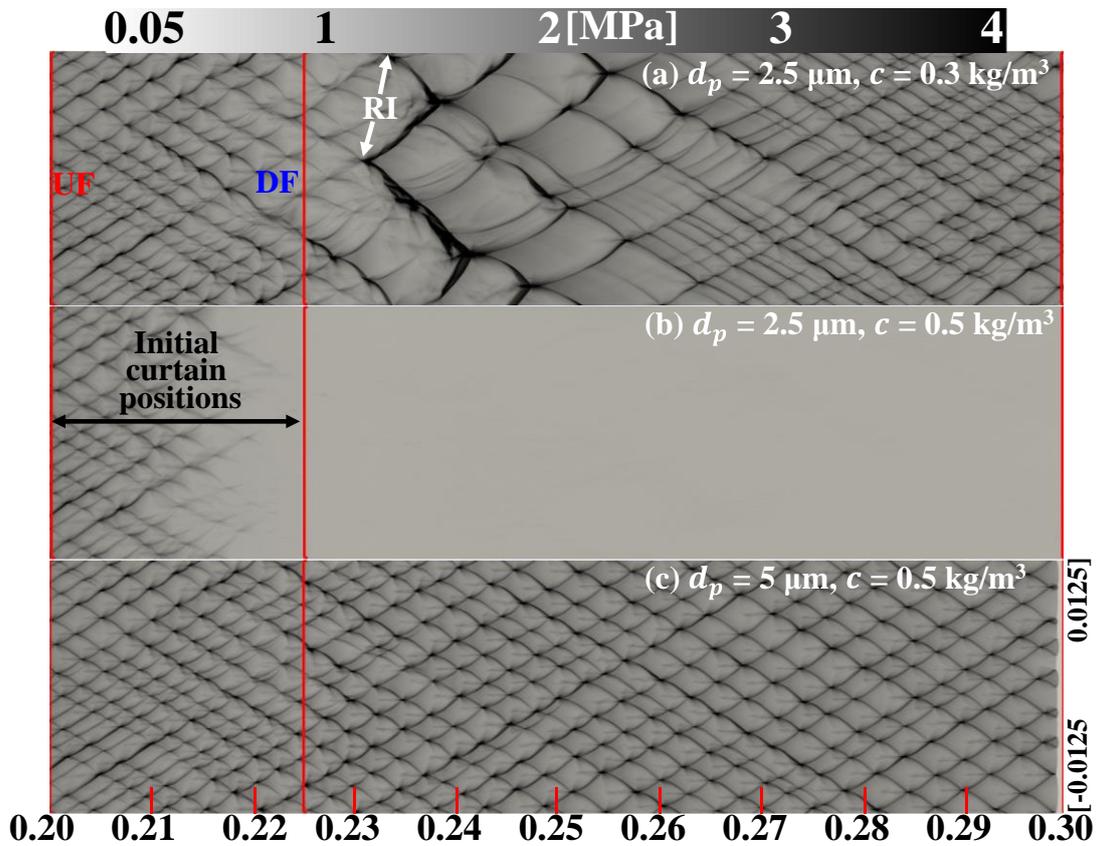

Figure 15 Peak pressure trajectories with various particle sizes and concentrations. The red lines denote the initial curtain locations. UF/DF: upstream/downstream front. The curtain thickness is $L$ = 25 mm. Axis label unit: m.

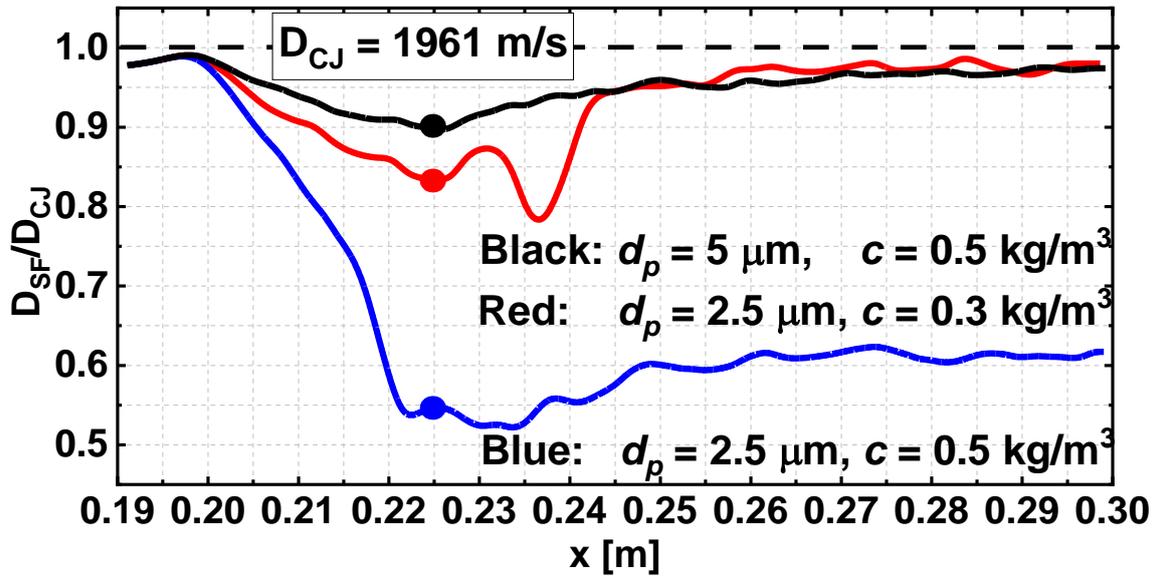

Figure 16 Effects of particle size and concentration on the shock speed. $D_{CJ}$ = 1961 m/s. Solid symbols indicate the downstream front of the particle curtain

In addition, denser curtain in Fig. 15(b) has longer transmission time, 18 μs, than the dilute cases (e.g., 16 μs in Fig. 15a) due to lower shock speeds in the transmission failure case. The extinction distance of Fig. 15(b) is about 22 mm, much shorter than 32 mm in Fig.11. Here we define the extinction distance based on the width-averaged peak pressure: when the average peak pressure is lower than 0.8 MPa, which is deemed as detonation extinction. This demonstrates that smaller particles are more effective in mitigating detonation (Fedorov & Kratova 2015). Furthermore, we increase the particle size from 2.5 μm in Fig. 15(b) to 5 μm in Fig. 15(c) at the same concentration. The results show that the detonation becomes transmissible with regular cell size in Fig. 15(c). This is because the smaller particles have larger momentum exchange rate due to shorter relaxation timescale, and also higher heat transfer rate resulting from greater specific surface area. The results in Fig. 15(c) can be further parameterized with the evolution of $D_{SF}$ (black line in Fig. 16). The shock speed is slightly reduced to around $0.9D_{CJ}$ at around 0.225 m (i.e., DF of the curtain). Subsequently, it is gradually recovered to the initial value, $0.98D_{CJ}$, at the curtain leeward. The higher shock speed leads to shorter transmission time in this case, i.e., about 15 μs.

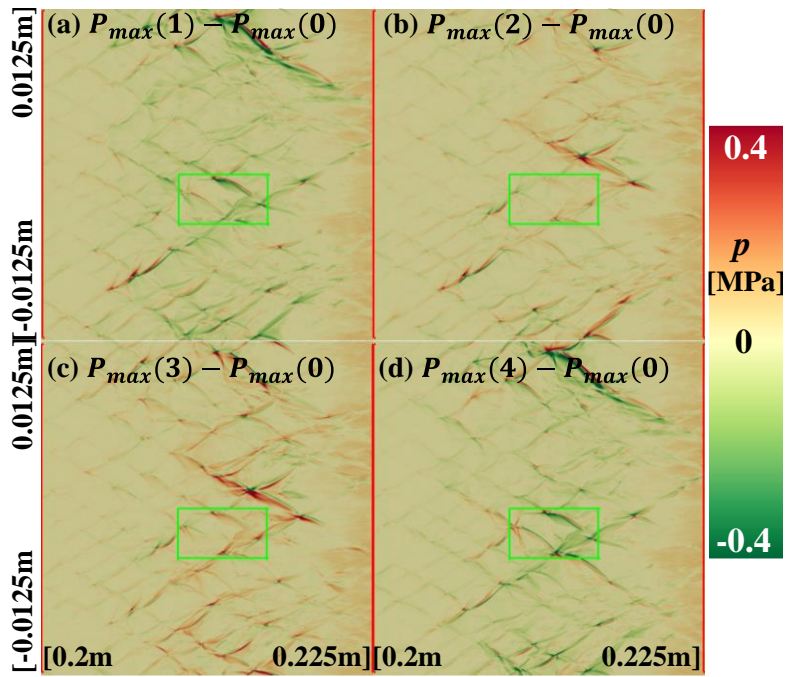

Figure 17 Peak pressure difference with various particle sizes and concentrations. The number in the brackets, e.g., $P_{max}(0)$, is the case number, and the case information is tabulated in Table 1.

We further examine the effects of particle diameter and concentration through changing them by ±1%. Figure 17 shows the difference of the peak pressure in the curtain between different cases, and the corresponding case information is presented in Table 1. Case 0 has the same curtain conditions as that of Fig. 15 (b). Compared with case 0, particle concentration in cases 1/2 increases / deceases by 1%, whilst they have the same diameter to case 0. However, for cases 3/4 with the same concentration, their diameters have become 101% / 99% of the initial value. In all cases, the detonation is decoupled, and no re-initiation is observed. However, it can be seen from Fig. 17 that small difference of particle size and concentration could considerably change the trajectories of peak pressure. This can be more clearly seen from Fig. 18 about the enlarged views of a single detonation cell in the boxes in Fig. 17. The dashed lines in Fig. 18 are the tracks of the triple points in case 0. All triple point tracks are generally coincident with each other, as also demonstrated with Section B of the supplementary document. In Fig.18, red/green area corresponds to increased/reduced peak pressure, particularly, in the second half of the detonation cell. For green area in Figs. 18(a) and18(d), overpressure reduction of the triple points and the incident wave is induced by the increased momentum and heat exchanges between the leading front and shocked particles. However, increased peak pressure can be seen at red area in Figs. 18(b) and 8(c), and this is because the local gas pressure is stronger due to smaller particle concentration or larger diameter.

Table 1 Numerical experiments with perturbed particle diameters and concentrations

| Case | Diameter [μm] | Concentration [g/m$^3$] |
|---|---|---|
| 0 | 2.5 | 500 |
| 1 | 2.5 | 505 |
| 2 | 2.5 | 495 |
| 3 | 2.525 | 500 |
| 4 | 2.475 | 500 |

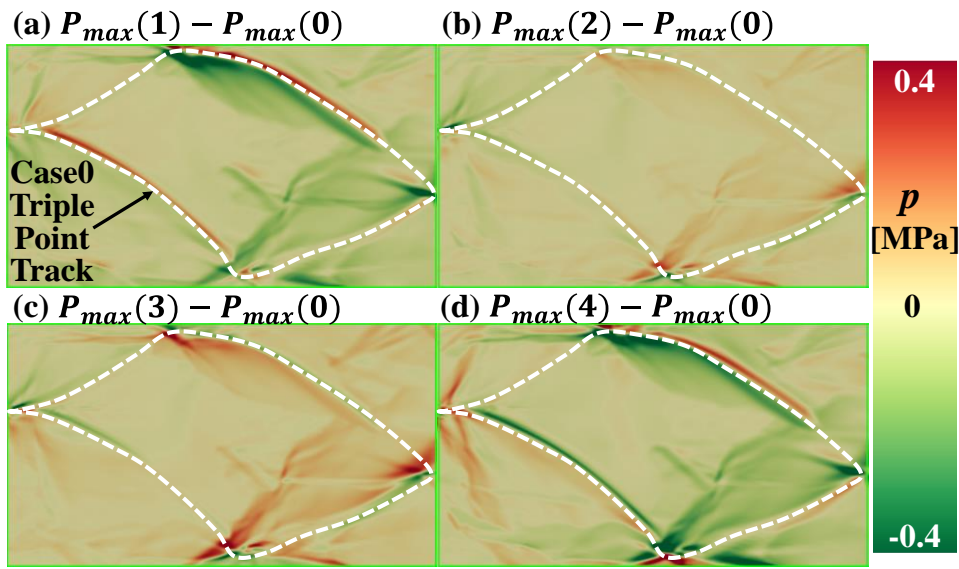

Figure 18 Peak pressure difference within a single detonation cell from the boxes in Fig. 17. Cases are tabulated in Table 1. Implication for the number in the brackets same as in Fig. 17.

**4.2.5 Two-phase interaction**

The interphase transfer rates of energy and momentum in the cases from section 4.2.3 are plotted in Fig. 19. The $S_{mom}$ and $S_{energy}$ (see Eqs. 13 and 14) are volume averaged based on the domain ($x = 0 – 0.3$ m). A negative value indicates that the energy or momentum is transferred from the gas to particles. One can see from Fig. 19 that the magnitudes of both energy and momentum transfer rates first increase and then decrease. The increase is induced by more particles in the post-detonation area as the DW travels in the curtain. Subsequently, the thermal and velocity equilibria between the gas and particles are achieved, and interphase exchanges gradually decrease. Beyond a critical time, both transfer rates approach zero, and this time increases with the curtain thickness. For instance, they are 40 and 85 μs for 10 and 50 mm curtains, respectively. It is observed that except the 10 mm curtain, all the others have the same peak values of energy/momentum transfer rate due to the same particle concentration. This is also probably associated with the particle momentum and thermal relaxation time because of the same particle size (5 μm) in these cases. In the 10 mm curtain, both transfer rates peak earlier than the rest since the equilibria cannot be reached within the shorter curtain.

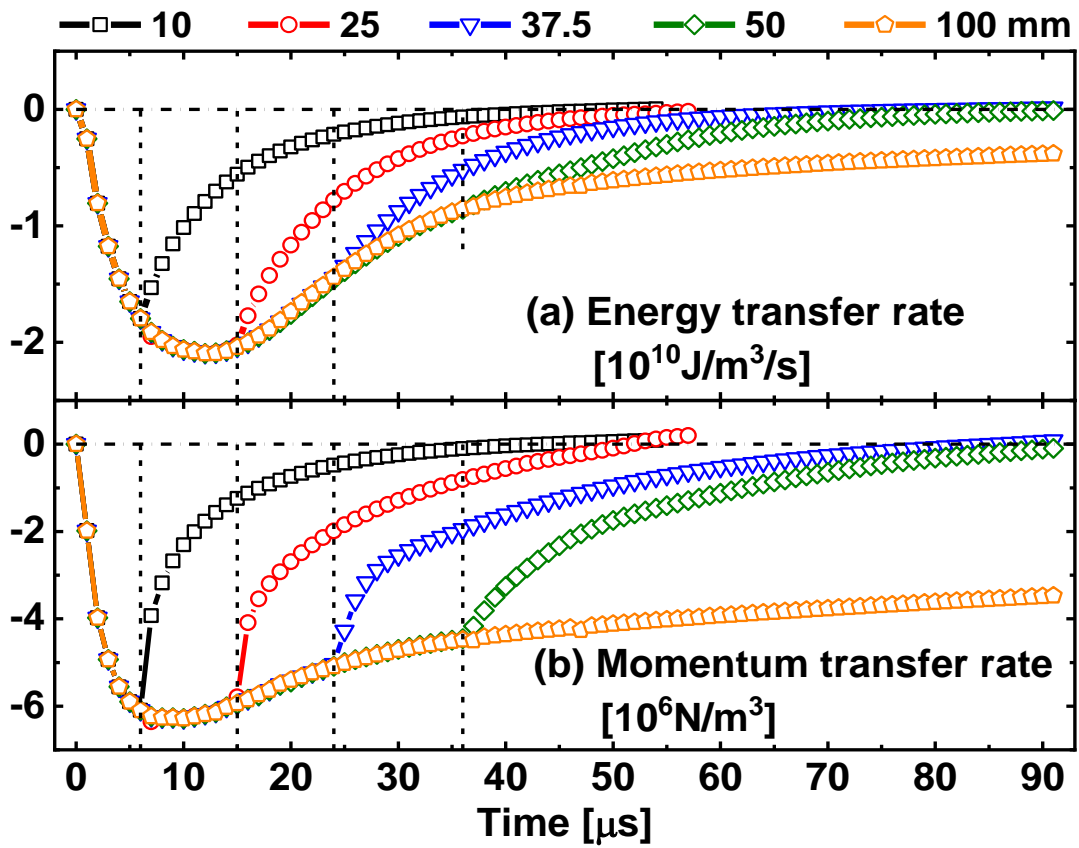

Figure 19 Time history of (a) energy and (b) momentum transfer rates with various curtain thicknesses. $c$ = 0.75 kg/m³ and $d_p$ = 5 μm. Vertical line: instant when the detonation or shock wave leave the curtain DF.

Likewise, the energy and momentum transfer rates studied for the particle concentration and size effects in Section 4.2.4 are presented with Fig. 20. Three curtain cases are used: case I: $d_p$ = 2.5 μm and $c$ = 0.3 kg/m³; case II: $d_p$ = 2.5 μm and $c$ = 0.5 kg/m³; case III: $d_p$ = 5 μm and $c$ = 0.5 kg/m³. Generally, similar evolutions of a transient increase and subsequent decrease can be seen from all cases.

Specifically, concentration effects from cases I and II show that denser particle curtain (case II) exhibits faster interphase transfer of energy and momentum, because of increased particle surface. Moreover, diameter effects from cases II and III demonstrate that the curtain with smaller particles (case II) is shown to have slightly greater magnitudes of interphase transfer rates due to greater particle specific surface area. Moreover, compared with the other results, longer residence time is observed from the case III due to its lower shock speed in the curtain. Note that convective heat transfer dominates interphase exchanges between the shock and particles, whilst momentum exchange is secondary, as shown in Section C of the supplementary document.

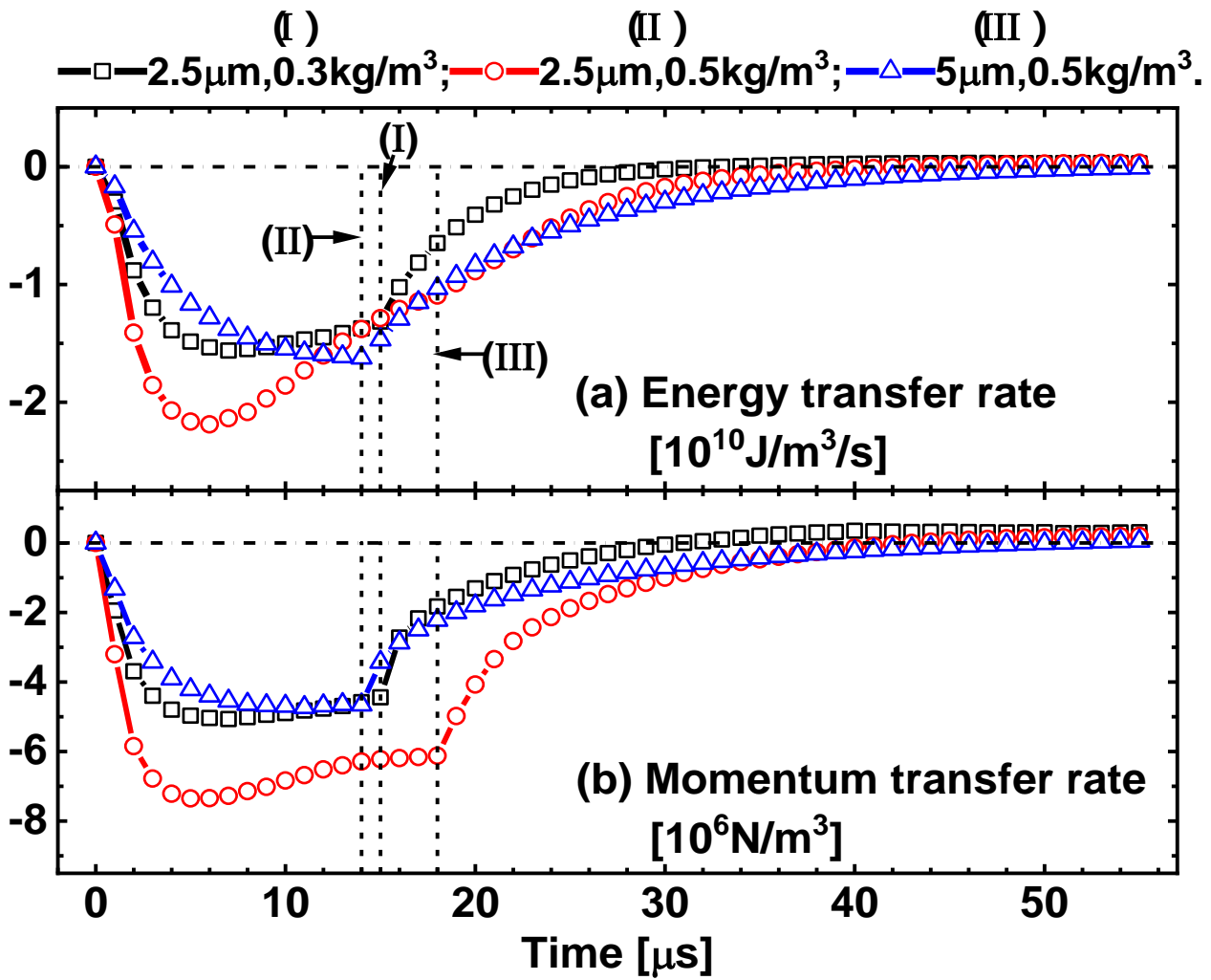

Figure 20 Time history of (a) energy and (b) momentum transfer rates with various particle concentrations and diameters. $L = 25$ mm. Vertical lines: residence time.

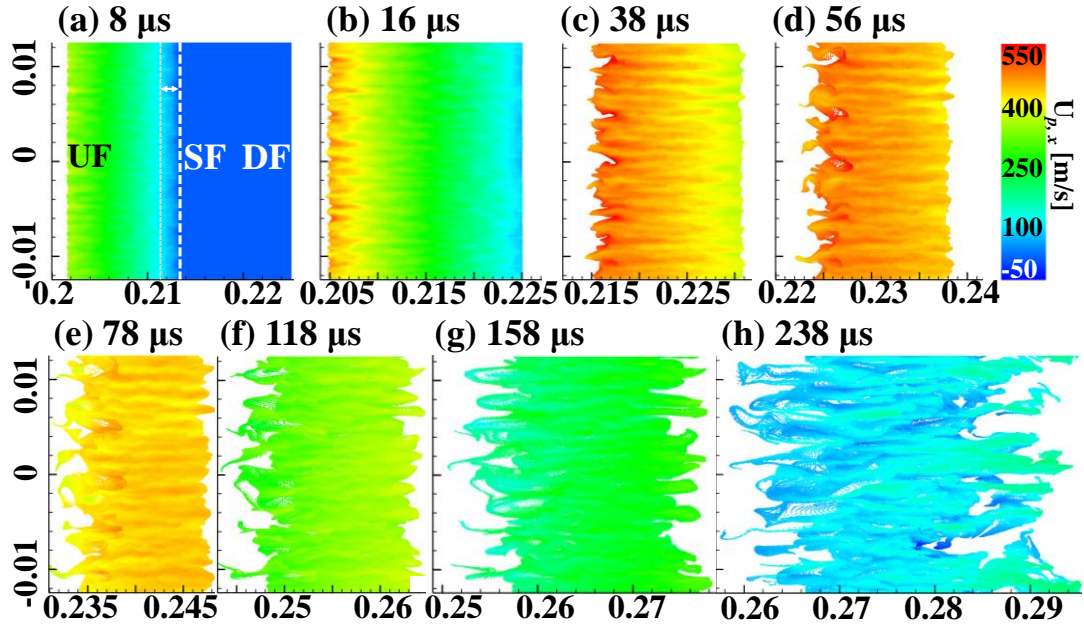

Figure 21 Evolution of curtain morphologies colored with particle *x*-direction velocity. $d_p$ = 5 μm, $c$ = 0.75 kg/m³, and $L$ = 25 mm. SF: shock front; UF/DF: upstream/downstream front. Axial label unit: m.

**4.3 Evolution of curtain morphology after detonation impacting**

Up to this point, we have studied hydrogen detonation dynamics subject to an inert particle curtain. In this section, we will further discuss how the shock affects the morphologies of the particle curtain. The evolutions of curtain morphologies are shown in Fig. 21, colored by the *x*-direction particle velocities. The one-dimensionalized (averaged along the domain width) profiles of particle volume fraction, *x*-direction velocity, temperature, and gas pressure are plotted in Fig. 22. The curtain conditions are $d_p$ = 5 μm, $c$ = 0.75 kg/m³, and $L$ = 25 mm. At 8 μs, the particle velocity and temperature gradually increase behind the leading SF, as shown in Fig. 21(a) and 22(a). This is due to the finitely long momentum and thermal response time of solid particles. At this instant, the particles at the curtain UF are accelerated to over 360 m/s, but those at the DF are still intact since the shock has not arrived yet. Due to the movement of UF particles, the 25 mm curtain shrinks to 23 mm at 8 μs.

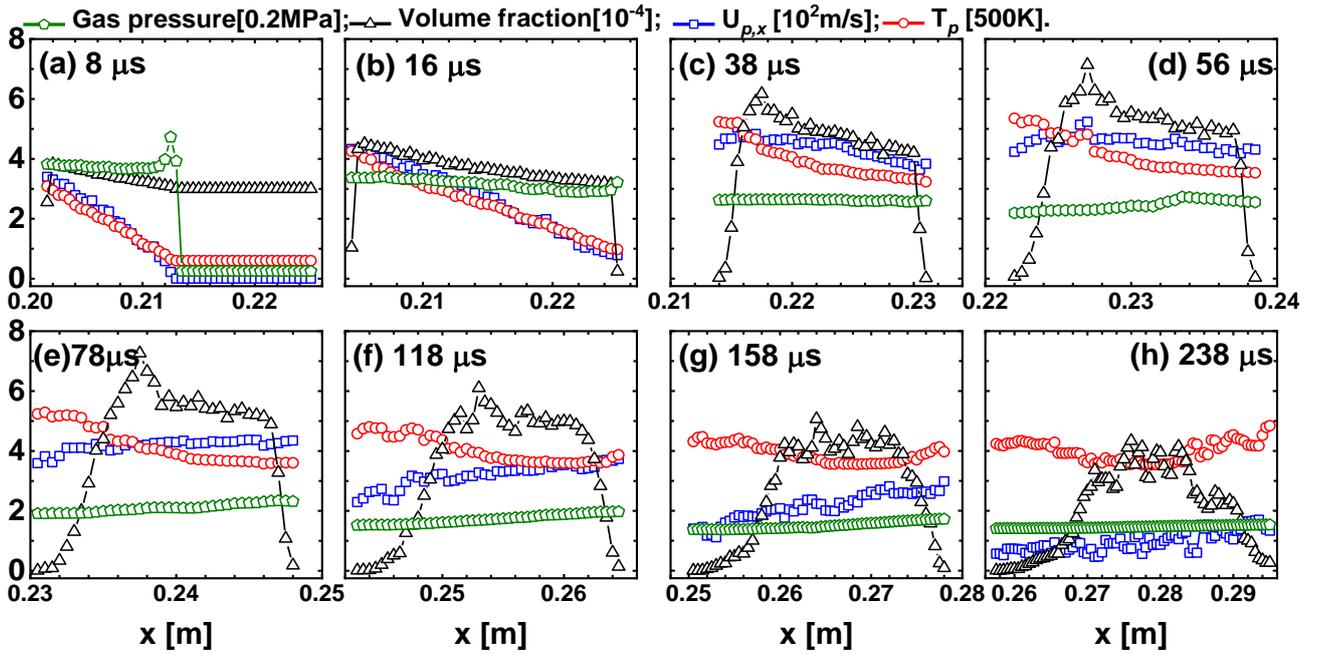

Figure 22 Evolutions of particle volume fraction, $x$-direction velocity, temperature, and gas pressure. $d_p = 5$ μm, $c = 0.75$ kg/m$^3$, and $L = 25$ mm.

The DW leaves the curtain at 16 μs, but the curtain continues changing due to the quickly evolving detonated flows from 38 to 78 μs as demonstrated in Figs. 21 and 22. During this period, spatial distributions of particle temperature, velocity, and curtain thickness (around 18 mm) do not change as pronouncedly as at 8 and 16 μs. This can be justified by the fact that now the entire curtain has been shocked and meanwhile almost all particles have already responded to the local flows to some degrees (i.e., being heated and moved). Meanwhile, one peak of particle volume fraction is observed near the curtain UF, indicating the local particle accumulation. This is because the particle number behind the SF is increased due to the Stokes drag force acting on these particles (see Section D of the supplementary document). At the last three instants in Figs. 21 and 22, i.e., 118 to 238 μs, the curtain thickness is continuously increased, accompanied with significantly reduced particle velocity in Figs. 22(e)─22(f). This is somehow affected by the quickly reduced gas velocity behind the detonation (Xu, Zhao & Zhang 2021; Watanabe et al. 2020). However, only small change is observed for particle temperature, because of the fast heat exchange with the hot detonation gas.

It must be mentioned that in Fig. 22 the pressure distributions across the curtain are relatively uniform for all instants, different from the dense curtain results by Theofanous & Chang (2017). Based on their results, pronounced pressure gradient are observed between the UF and DF of the curtain, which drives the evolutions of curtain morphology, including constant acceleration regime (CAR) and constant velocity regime (CVR). However, since dilute curtain (initial volume fraction 0.03%) is considered in our study, the acoustic impedance is relatively small (about 256 kg/m$^2$/s) across the UF/DF of the curtain. As mentioned above, no reflected shock or rarefaction wave is generated from them to induce the pressure discontinuity and affect the pressure distribution in the curtain.

Since pressure gradient is almost absent across the dilute curtain but we still observe the ostensibly similar morphological change to that of the dense one by Theofanous & Chang (2017), it is necessary to further explore the reasons from our results. Figure 23 shows the evolutions of the drag and Archimedes force acting on the particles at the instants in Figs. 21 and 22. Note that a positive force indicates that it is aligned with the direction of detonation wave propagation. Apparently, the drag force $\mathbf{F}_{d,p}$ is always approximately 1-2 orders of magnitudes higher than the pressure gradient force $\mathbf{F}_{p,p}$, indicating the importance of $\mathbf{F}_{d,p}$ in influencing the change of curtain morphologies. In Fig. 23(a), the magnitudes of $\mathbf{F}_{d,p}$ and $\mathbf{F}_{p,p}$ peak immediately behind the SF. Thereafter, in Fig. 23(b), their magnitudes are reduced. This is because the shock wave is weakened in the curtain through energy and momentum extraction. Meanwhile, the particle acceleration, to some degree, also minimizes the interphase velocity difference (hence reduced drag force). From Fig. 23(c) to 23(e), negative drag force $\mathbf{F}_{d,p}$ occurs near the UF due to larger particle speeds than those of the gas phase in the post-detonation zone. Take a typical instant, 56 μs in Fig. 23(d), as an example. It is shown that the negative drag forces are comparable (in magnitude) to the positive counterpart, which results in limited change of the curtain thickness (about 18 mm) over this period. In last three instants, almost all the drag force becomes negative, indicating that they are in the opposite direction of curtain movement and overall speed of curtain speed may be reduced. Nonetheless, the particle velocities at the UF are generally lower than those at the DF (see the profiles at 158 μs from Fig. 23g), and the curtain thickness becomes larger as revealed in Fig. 21.

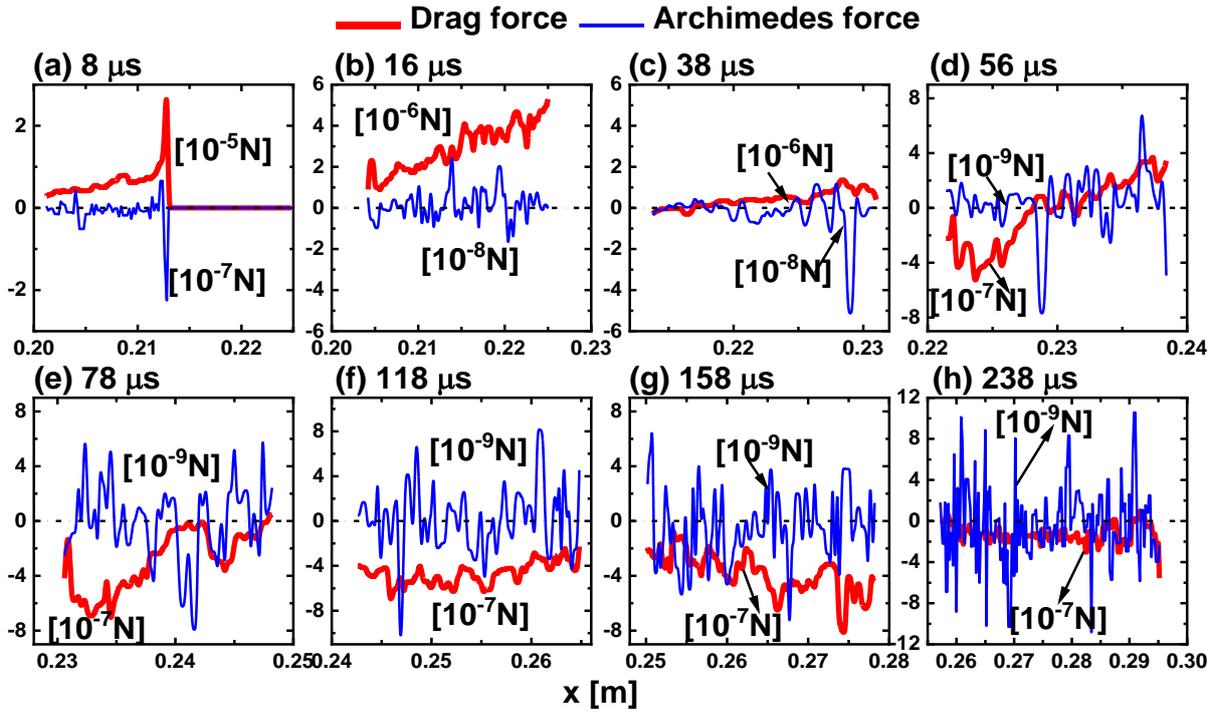

Figure 23 Time history of drag force and Archimedes force acting on the particles. $d_p = 5$ μm, $c = 0.75$ kg/m$^3$, and $L = 25$ mm.

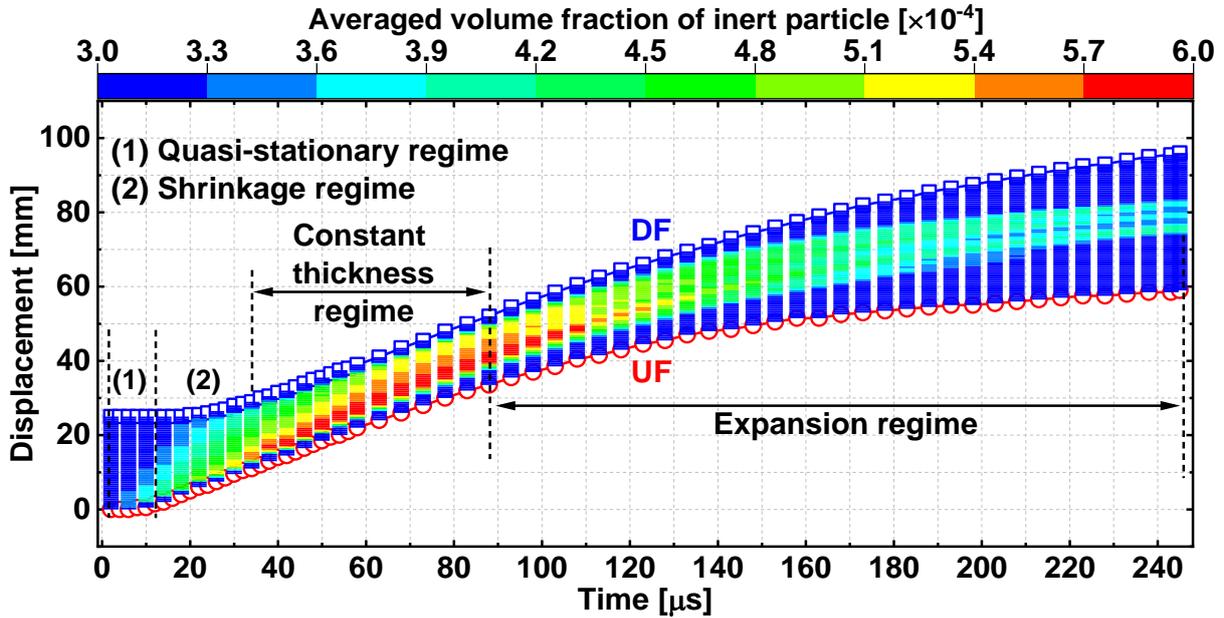

Figure 24 Trajectories of UF/DF, and evolutions of averaged particle volume fraction inside the domain ($x = 0.2 - 0.3$ m). $d_p = 5$ μm, $c = 0.75$ kg/m$^3$, and $L = 25$ mm. From 0 μs to 58 μs, the detonation is inside the curtain.

The displacements of curtain UF/DF and evolutions of particle distribution in the curtain are shown in Fig. 24. Note that the particle volume fraction for each instant is averaged along the width of the computational domain. The UF/DF displacements are recorded relative to the UF of the initial curtain (i.e., $x = 0.2$ m), and the streamwise positions of UF/DF are identified from the location where the width-averaged volume fraction is critically above 1% of the initial value. The curtain thickness therefore can be estimated based on the difference between the UF and DF locations. Various regimes of curtain evolution are identified: *quasi-stationary regime* (QSR), *shrinkage regime* (SR), *constant-thickness regime* (CTR), and *curtain expansion regime* (CER). The quantitative characterization of the four curtain regimes is detailed in Fig. 25 and Table 2. For each regime, the evolutions of the curtain thickness with time are quantified with the dimensionless time $t^* \equiv t \times (D_{SF}/L_0)$ and the corresponding fitted functions for each regime are tabulated in Table 2. Here $D_{SF}$ is the initial shock speed and $L_0$ is the initial curtain thickness. In this case, they are 1,961 m/s and 25 mm, respectively. $L^*$ is the dimensionless curtain thickness, i.e., $L^* \equiv L/L_0$. The relations between $L^*$ and $t^*$ are plotted in Fig. 25 and tabulated in Table 2.

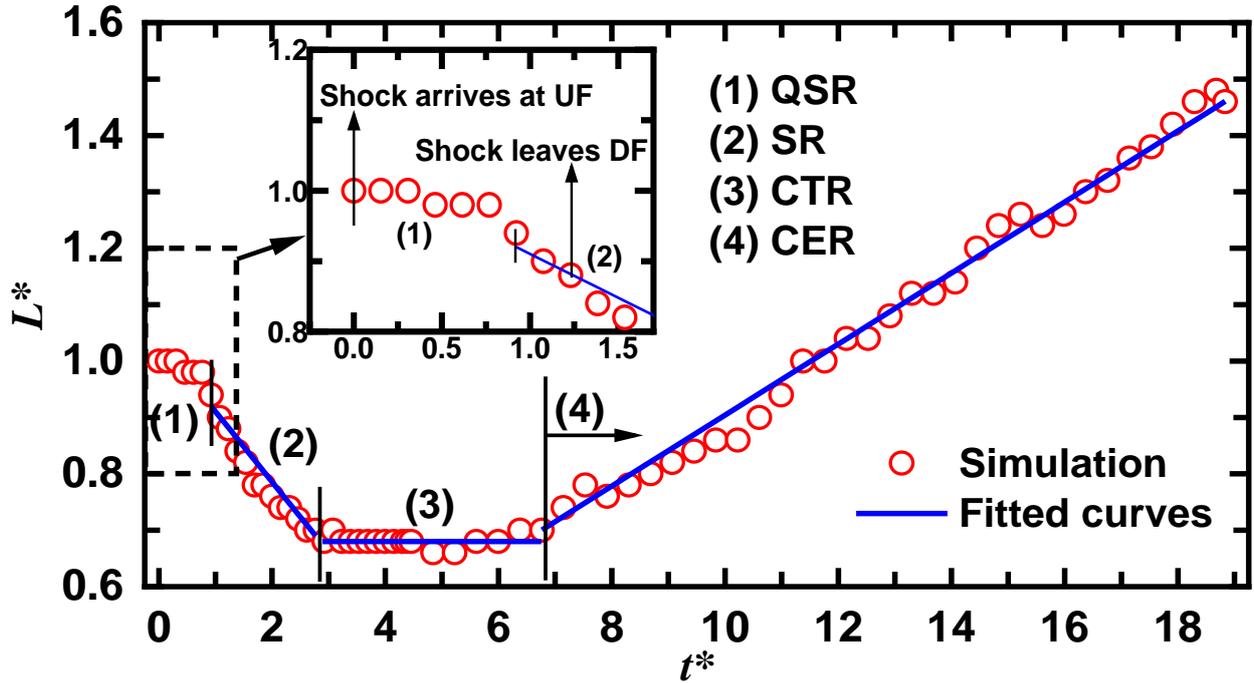

Figure 25 Curtain evolution regimes. $d_p = 5$ μm, $c = 0.75$ kg/m³, and $L = 25$ mm.

Table 2 Fitted functions of curtain thickness in various curtain regimes

| Regime | Curtain thickness equation | Duration | Eq. |
|---|---|---|---|
| Quasi-stationary regime (QSR) | $L^* = 0.92 \sim 1$, | $0 < t^* < 0.92$ | (15) |
| Shrinkage regime (SR) | $L^* = 0.92 - 0.045(t^* - 0.92)$, | $0.92 < t^* < 2.89$ | (16) |
| Constant-thickness regime (CTR) | $L^* = 0.68 \sim 0.7$, | $2.89 < t^* < 6.75$ | (17) |
| Curtain expansion regime (CER) | $L^* = 0.7 + 0.02275(t^* - 6.75)$, | $t^* > 6.75$ | (18) |

Initially ($t^* < 0.92$ in Table 2), a *quasi-stationary regime* (QSR) can be seen in Fig.25 and Table 2, in which the curtain thickness does not change. This is because the particles at the UF take a short period to respond to the shock. Thereafter, due to shock impacting, these particles at the curtain UF start to move, driven by the drag force, leading to curtain compression. This regime spans from $t^* = 0.92$ to about $t^* = 2.89$ when the UF and DF begin to have the same speed (see Fig. 25). It is the *shrinkage regime* (SR), in which the dimensionless curtain thickness $L^*$ is linearly reduced with $t^*$, see Eq. (16). The curtain thickness is reduced from 24 mm at $t^* = 0.92$ to 18 mm at $t^* = 2.89$. After that, the curtain thickness has negligible change, which is *constant-thickness regime* (CTR). This regime lasts from $t^* = 2.89$ to $t^* = 6.75$. However, as can be seen from Fig. 24, the particles are continuously re-distributed, leading to increased volume fraction in the curtain interior. Subsequently, a *curtain expansion regime* (CER) is observed, and the expansion rate of the curtain can be described with $L^* = 0.7 + 0.02275(t^* - 6.75)$.

Now we further discuss the difference and similarity between the Theofanous et al. experimental results and our simulations regarding regimes of curtain morphology. In their dilute curtain experiments (Theofanous et al. 2018), only expansion regime is observed, including accelerated and constant expansion. In fact, a weak decrease of the curtain thickness is also present around 0.3 ms after shock arrival, but they did not delve into it probably due to insufficient time resolution to capture this process. Similar expansion regimes were also observed in their dense curtain experiments (Theofanous et al. 2016). However, when the particle concentration is high, the curtain expansion with a constant acceleration happens immediately, followed by linear expansion period. Their dense curtain expansion behaviors almost occupy entire process, which is different from our dilute curtain expansion that occurs at the latter half of evolutions. This is because both Archimedes force and drag force are significant in inducing dense curtain movement.

The above regimes also occur in 10 mm and 37.5 mm curtains, and the reader can refer to Section E of the supplementary document for details. Nonetheless, the particle distributions experience different evolutions, e.g., from single peak to multiple peaks, which may be related to various distributions of particle forces. The influences of curtain thickness on the characteristic time of curtain evolutions are further examined in Fig. 26. We can see that the shock arrives ($t_{in}$) or leaves ($t_{out}$) the curtian almost at the same dimensionless time, which is because of similar incident shock speeds. Moreover, the time instants of onset of the SR, CTR and CER (i.e., $t_{SR}$, $t_{CTR}$, and $t_{CER}$, as annotated in Fig. 26) generally become smaller, when the curtain become thicker. This trend is justified by the fact that the drag force acting on the particles generally become greater due to increased particle number.

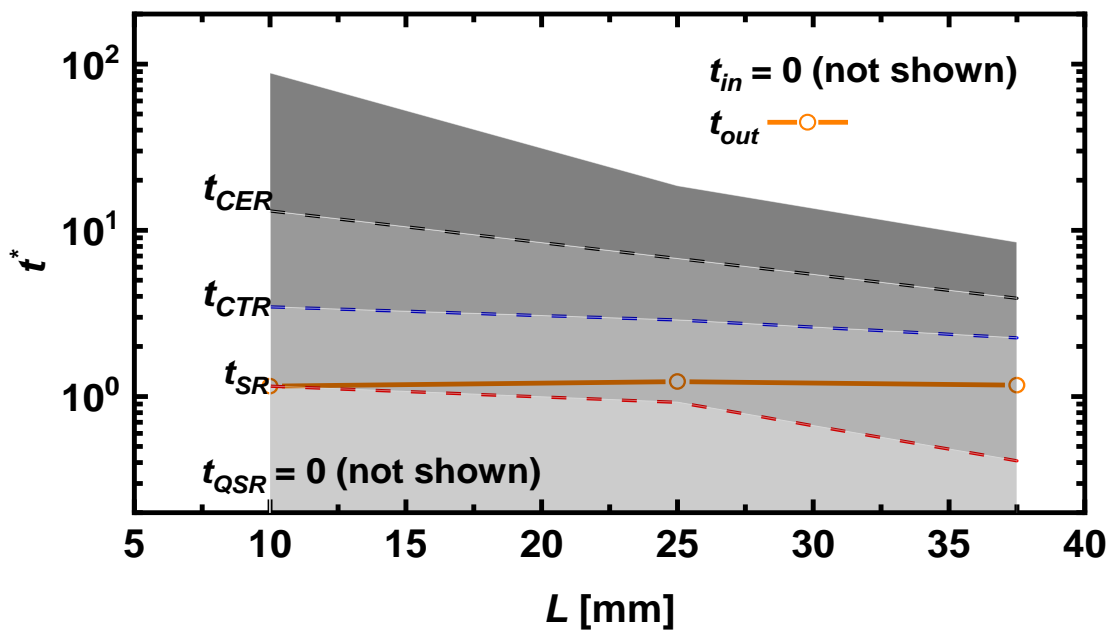

Figure 26 Characteristic time of curtain evolutions with various curtain thickness. $d_p$ = 5 μm and $c$ = 0.75 kg/m³.

## 5. Conclusions

The stoichiometric hydrogen/air detonation propagation in an inert particle curtain is simulated by the Eulerian─Lagrangian method with gas-particle two-way coupling. A map of critical curtain thickness involving various particle sizes, concentrations and curtain thicknesses is compiled through parametric studies. The critical curtain thickness for detonation extinction reduces quickly with the particle concentration for the three particle diameters studied. Beyond the critical particle concentration, the critical curtain thickness exhibits weak dependence on concentration. Moreover, for a fixed curtain thickness, with increased particle diameter, the concentration range to achieve extinction is appreciably elevated.

The gas-particle detonation frontal structures are investigated and it is found that the thermal and velocity equilibria are not reached as the shock passes through the curtain. Unsteady detonation phenomena, such as detonation extinction and re-initiation, are analyzed with shock frontal structures, chemical explosive mode, and shock wave structure. It is demonstrated that the detonation can be re-initiated by a hot point resulting from transverse shock focusing.

The particle curtain thickness has significant effects on peak pressure trajectory, shock speed, and heat release rate. Three propagation modes of the detonation wave in particle curtain are identified: (1) detonation transmission, (2) partial extinction and detonation re-initiation, and (3) detonation extinction. Numerical experiments confirm that the occurrence of a re-initiation point is not a fully stochastic event, but the location is very sensitive to the curtain properties. Furthermore, the influence of particle diameter/concentration on the DW is also examined with peak pressure trajectories, shock speeds, and interphase exchange rates of energy and momentum. Meanwhile, numerical experiments for two-phase interactions show the key role of interphase heat exchanges in detonation inhibition.

The evolutions of curtain morphologies are clarified, through the particle velocity and volume fraction, and Stokes drag and Archimedes forces. Due to the Stokes drag force, different evolution regimes are formed: quasi-stationary regime, shrinkage regime, constant-thickness regime, and expansion regime. Moreover, the influence of the curtain thickness on the characteristic time of curtain evolutions is also discussed.

**Acknowledgements**

This work used the Fugaku Supercomputer in RIKEN Center for Computational Science, Japan (HP 210196). YX is supported by the NUS Research Scholarship. This work is finally supported by Singapore Ministry of Education Tier 1 Grant (A-0005238-01-00).

**Declaration of interests**

The authors report no conflict of interest.

**References**

BURKE, M. P., CHAOS, M., JU, Y., DRYER, F. L. & KLIPPENSTEIN, S. J. 2012 Comprehensive $H_2/O_2$ kinetic model for high-pressure combustion. *Int. J. Chem. Kinet.* **44**, 444–474.
BOIKO, V. M., KISELEV, V. P., KISELEV, S. P., PAPYRIN, A. N., POPLAVSKY, S. V. & FOMIN, V. M. 1997 Shock wave interaction with a cloud of particles. *Shock Waves* **7**, 275–285.
BHATTACHARJEE, R. R., LAU-CHAPDELAINE, S. S. M., MAINES, G. MALEY, L. & RADULESCU, M. I. 2013 Detonation re-initiation mechanism following the Mach reflection of a quenched detonation. *Proc. Combust. Inst.* **34**, 1893–1901.
BOIKO, V. M., KISELEV, V. P., KISELEV, S. P., PAPYRIN, A. N., POPLAVSKII, S. V. & FOMIN, V. M. 1996 Interaction of a shock wave with a cloud of particles. *Combust. Explos. Shock Waves* **32**, 191–203.
CROWE, C. SCHWARZKOPF, J. SOMMERFELD, M. & TSUJI, Y. 2011 Multiphase Flows with Droplets and Particles, 2nd ed., CRC Press.
CROWE, C. T. SHARMA, M. P. & STOCK, D. E. 1977 The particle-source-in cell (PSI-CELL) model for gas-droplet flows. *J. Fluids Eng.* **99**, 325–332.
FEDOROV, A. V., TROPIN, D. A. & BEDAREV, I. A. 2010 Mathematical modeling of detonation suppression in a hydrogen-oxygen mixture by inert particles. *Combust. Explos. Shock Waves* **46**, 332–343.
FEDOROV, A. V. & KRATOVAA, Y. V. 2015a Influence of non-reactive particle cloud on heterogeneous detonation propagation. *J. Loss Prev. Process Ind.* **36**, 404–415.
FEDOROV, A. V. & KRATOVAA, Y. V. 2015b Analysis of the influence of inert particles on the propagation of a cellular heterogeneous detonation. *Shock Waves* **25**, 255–265.
FEDOROV, A. V. & TROPIN, D. A. 2013 Modeling of detonation wave propagation through a cloud of particles in a two-velocity two-temperature formulation. Combust. Explos. *Shock Waves* **49**, 178–187.
FEDOROV, A. V. & TROPIN, D.A. 2011 Determination of the critical size of a particle cloud necessary for suppression of gas detonation. *Combust. Explos. Shock Waves* **47**, 464–472.
FEDOROV, A. V. & KRATOVAA, Y. V. 2013 Calculation of detonation wave propagation in a gas suspension of aluminum and inert particles. *Combust. Explos. Shock Waves* **49**, 335–347.
FOMIN, P. A. & CHEN, J. R. 2009 Effect of chemically inert particles on thermodynamic characteristics and detonation of a combustible gas. *Combust. Sci. Technol.* **181**, 1038–1064.
GOTTIPARTHI, K.C. & MENON, S. 2012 A study of interaction of clouds of inert particles with detonation in gases. *Combust. Sci. Technol.* **184**, 406–433.

HUANG, Z. ZHAO, M. XU, Y. LI, G. & ZHANG, H. 2021 Eulerian-Lagrangian modelling of detonative combustion in two-phase gas-droplet mixtures with OpenFOAM: Validations and verifications. *Fuel*. **286**, 119402.

KHMEL, T. A. & FEDOROV, A. V. 2014 Modeling of propagation of shock and detonation waves in dusty media with allowance for particle collisions. *Combust. Explos. Shock Waves* **50**, 547–555.

KRATOVAA, Y. V. & FEDOROV, A. V. 2014 Interaction of a heterogeneous detonation wave propagating in a cellular regime with a cloud of inert particles. *Combust. Explos. Shock Waves* **50**, 183–191.

KURGANOV, A., NOELLE, S. & PETROVA, G. 2001 Semidiscrete central-upwind schemes for hyperbolic conservation laws and Hamilton-Jacobi equations. *SIAM J. Sci. Comput.* **23**, 707–740.

LIU, Q., HU, Y., BAI, C. & CHEN, M. 2013 Methane/coal dust/air explosions and their suppression by solid particle suppressing agents in a large-scale experimental tube. *J. Loss Prev. Process Ind.* **26**, 310–316.

LIU, Y., LIU, X. & LI, X. 2016 Numerical investigation of hydrogen detonation suppression with inert particle in pipelines. *Int. J. Hydrogen Energy* **41**, 21548–21563.

LIU, A. B. MATHER, D. & REITZ, R.D. 1993 Modeling the effects of drop drag and breakup on fuel sprays. *SAE Tech. Pap.* 83–95.

LU, T. F. YOO, C. S. CHEN, J. H. & LAW, C. K. 2010 Three-dimensional direct numerical simulation of a turbulent lifted hydrogen jet flame in heated coflow: A chemical explosive mode analysis. *J. Fluid Mech.* **652**, 45–64.

LAM, S. H. 1985 Singular Perturbation for Stiff Equations Using Numerical Methods. *In Recent advances in the aerospace sciences*. Springer, Boston, MA. 3–19.

LAM, S. H. & GOUSSIS, D. A. 1994 The CSP method for simplifying kinetics. *Int. J. Chem. Kinet.* **26**, 461–486.

LAM, S. H. 1993 Using CSP to Understand Complex Chemical Kinetics. *Combust. Sci. Technol.* **89**, 375–404.

LAM, S.H. 2007 Reduced chemistry-diffusion coupling. *Combust. Sci. Technol.* **179**, 767–786.

MENG, Q., ZHAO, M., ZHENG, H. & ZHANG, H. 2020 Eulerian − Lagrangian modelling of rotating detonative combustion in partially pre-vaporized n-heptane sprays with hydrogen addition. *Fuel* **290**, 119808.

OLMOS, F. & MANOUSIOUTHAKIS, V. I. 2013 Hydrogen car fill-up process modeling and simulation. *Int. J. Hydrogen Energy* **38**, 3401–3418.

PINAEV, A. V. VASILEV, A.A. & PINAEV, P. A. 2015 Suppression of gas detonation by a dust cloud at reduced mixture pressures. *Shock Waves* **25**, 267–275.

RANJAN, D., OAKLEY, J. & BONAZZA, R. 2011 Shock-bubble interactions. *Annu. Rev. Fluid Mech.* **43**, 117–140

RANZ, W. E. & MARSHALL, W. R. 1952 Evaporation from Drops, Part I. *Chem. Eng. Prog.* **48**, 141–146.

SHI, J., XU, Y., REN, W. & ZHANG, H. 2022 Critical condition and transient evolution of methane detonation extinction by fine water droplet curtains. *Fuel* **315**, 123133.

SONG, Y., NASSIM, B. & ZHANG, Q. 2018 Explosion energy of methane/deposited coal dust and inert effects of rock dust. *Fuel* **228**, 112–122.

TAHSINI, A.M. 2016 Detonation wave attenuation in dust-free and dusty air. *J. Loss Prev. Process Ind.* **39**, 24–29.

TROPIN, D.A. & FEDOROV, A. V. 2018 Attenuation and Suppression of Detonation Waves in Reacting Gas Mixtures by Clouds of Inert Micro- and Nanoparticles. *Combust. Explos. Shock Waves* **54**, 200–206.

TROPIN, D. & FEDOROV, A. 2019 Physical and mathematical modeling of interaction of detonation waves in mixtures of hydrogen, methane, silane, and oxidizer with clouds of inert micro-and nanoparticles. *Combust. Sci. Technol.* **191**, 275–283.

TROPIN, D.A. & FEDOROV, A. V. 2014 Mathematical modeling of detonation wave suppression by cloud of chemically inert solid particles. *Combust. Sci. Technol.* **186**, 1690–1698.

THEOFANOUS, T. G. MITKIN, V. & CHANG, C. H. 2016 The dynamics of dense particle clouds subjected to shock waves. Part 1. Experiments and scaling laws. *J. Fluid Mech.* **792**, 658–681.
THEOFANOUS, T. G. MITKIN, V. & CHANG, C. H. 2018 Shock dispersal of dilute particle clouds. *J. Fluid Mech.* **841**, 732-745.
TROPIN, D.A. & FEDOROV, A. V. 2014 Physicomathematical modeling of detonation suppression by inert particles in methane-oxygen and methane-hydrogen-oxygen mixtures. *Combust. Explos. Shock Waves* **50**, 542–546.
TROPIN, D. A. & BEDAREV, I. A. Problems of detonation wave suppression in hydrogen-air mixtures by clouds of inert particles in One- and Two-dimensional formulation. *Combust. Sci. Technol.* **193**, 197–210.
WATANABE, H. MATSUO, A. CHINNAYYA, A. MATSUOKA, K. KAWASAKI, A. & KASAHARA, J. 2020 Numerical analysis of the mean structure of gaseous detonation with dilute water spray. *J. Fluid Mech.* **887**, A4.
WELLER, H. G., TABOR, G., JASAK, H. & FUREBY, C. 1998 A tensorial approach to computational continuum mechanics using object-oriented techniques. *Comput. Phys.* **12**, 620.
WU, W. PIAO, Y. XIE, Q. & REN, Z. 2019 Flame diagnostics with a conservative representation of chemical explosive mode analysis. *AIAA J.* **57**, 1355–1363.
XU, Y. ZHAO, M. & ZHANG, H. 2021 Extinction of incident hydrogen/air detonation in fine water sprays. *Phys. Fluids* **33**, 116109.
XU, Y. & ZHANG, H. 2022 Pulsating propagation and extinction of hydrogen detonations in ultrafine water sprays. *Combust. Flame* **241**, 112086.
ZHAO, M., REN, Z. & ZHANG, H. 2021 Pulsating detonative combustion in n-heptane/air mixtures under off-stoichiometric conditions. *Combust. Flame* **226**, 285–301.
ZHAO, M., LI, J. M., TEO, C. J., KHOO, B. C. & ZHANG, H. 2020 Effects of variable total pressures on instability and extinction of rotating detonation combustion. *Flow, Turbul. Combust.* **104**, 261–290.
ZHAO, M. & ZHANG, H. 2020 Origin and chaotic propagation of multiple rotating detonation waves in hydrogen/air mixtures. *Fuel* **275**, 117986.
ZHAO, M. & ZHANG, H. 2020 Large eddy simulation of non-reacting flow and mixing fields in a rotating detonation engine. *Fuel* **280**, 118534.
ZHAO, M. & ZHANG, H. 2021 Rotating detonative combustion in partially pre-vaporized dilute n-heptane sprays: droplet size and equivalence ratio effects. *Fuel* (Accepted and In Press).
ZHAO, M., CHEN, Z. X., ZHANG, H. & SWAMINATHAN, N. 2021 Large Eddy simulation of a supersonic lifted hydrogen flame with perfectly stirred reactor model. *Combust. Flame* **230**, 0–51.
ZHOU, Z.Y. KUANG, S.B. CHU, K.W. & YU, A.B. 2010 Discrete particle simulation of particle-fluid flow: Model formulations and their applicability. *J. Fluid Mech.* **661**, 482–510.
ZHU, R., ZHAO, M. & ZHANG, H. 2021 Numerical simulation of flame acceleration and deflagration-to-detonation transition in ammonia-hydrogen–oxygen mixtures. *Int. J. Hydrogen Energy* **46**, 1273–1287.